\documentclass[authoryear,preprint,12pt]{elsarticle}
\usepackage{graphicx}
\usepackage{pdflscape}

\usepackage{booktabs}
\usepackage{tabularx}
\usepackage{array}
\usepackage{float}

\usepackage{tabularx}

\usepackage{algorithm}
\usepackage{algorithmic}
\usepackage{natbib}

\usepackage{color}
\newcommand{\red}[1]{#1}
\newcommand{\teal}[1]{#1}
\newcommand{\nv}[1]{#1}
\newcommand{\rv}[1]{#1}
\newcommand{\rw}[1]{#1}
\usepackage{graphicx}
\usepackage{lineno}
\usepackage{amssymb}
\usepackage{float}
\usepackage[section]{placeins} 

\usepackage{arydshln}
\usepackage{url}

\usepackage{multirow}
\usepackage{threeparttable}
\usepackage{bm}
\usepackage{amsmath}
\usepackage{booktabs} 
\usepackage{appendix}

\usepackage[x11names]{xcolor}
\usepackage{tikz}
\usetikzlibrary{arrows.meta, positioning, shapes.geometric, fit, calc, backgrounds}
\definecolor{modonecolor}{HTML}{6D2D50}
\definecolor{modtwocolor}{HTML}{7E88BE}
\definecolor{modthreecolor}{HTML}{7AA97A}
\definecolor{nvcolor}{HTML}{6F2DA8}
\newsavebox{\modOneSavebox}
\newsavebox{\modTwoSavebox}
\newsavebox{\modThreeSavebox}
\sbox{\modOneSavebox}{\tikz[baseline=-0.55ex]{\node[circle,fill=modonecolor,inner sep=0pt,minimum size=1.4em,text=white,font=\bfseries\small]{1};}}
\sbox{\modTwoSavebox}{\tikz[baseline=-0.55ex]{\node[circle,fill=modtwocolor,inner sep=0pt,minimum size=1.4em,text=white,font=\bfseries\small]{2};}}
\sbox{\modThreeSavebox}{\tikz[baseline=-0.55ex]{\node[circle,fill=modthreecolor,inner sep=0pt,minimum size=1.4em,text=white,font=\bfseries\small]{3};}}
\DeclareRobustCommand{\modOne}{\texorpdfstring{\usebox{\modOneSavebox}}{1}}
\DeclareRobustCommand{\modTwo}{\texorpdfstring{\usebox{\modTwoSavebox}}{2}}
\DeclareRobustCommand{\modThree}{\texorpdfstring{\usebox{\modThreeSavebox}}{3}}
\usepackage{listings}
\usepackage[
 colorlinks,
]{hyperref}

\AtBeginDocument{\hypersetup{citecolor=DodgerBlue4}}

\makeatletter
\def\ps@pprintTitle{%
 \let\@oddhead\@empty
 \let\@evenhead\@empty
 \def\@oddfoot{\footnotesize\itshape
 \hfill Accepted manuscript (in proof) --- Annals of the American Association of Geographers\hfill}%
 \let\@evenfoot\@oddfoot}
\makeatother

\journal{Annals of the American Association of Geographers}

\begin{document}

\begin{frontmatter}
\title{\nv{Toward Self-Evolving Geospatial Algorithms: An LLM-Driven Framework with a Case Study on Spatial Regime Discovery}\tnoteref{t1,t2}}


\tnotetext[t1]{This is the authors' accepted manuscript of an article accepted for publication in the \emph{Annals of the American Association of Geographers} (in proof). The final version may differ from this preprint.}
\tnotetext[t2]{Project page: \url{https://peng-luo.com/sega/}. It summarizes the framework and the results, and links to the code and data repository and to an interactive round-by-round trace of the evolution reported here.}

\author[a]{Xiayin Lou}
\ead{xiayin.lou@tum.de}

\author[b,c]{Peng Luo\corref{cor1}}
\cortext[cor1]{Corresponding author.}
\ead{peng-luo@uiowa.edu}

\author[a]{Liqiu Meng}
\ead{liqiu.meng@tum.de}

\address[a]{Chair of Cartography and Visual Analytics, Technical University of Munich, Munich, Germany}
\address[b]{School of Earth, Environment, and Sustainability, University of Iowa, Iowa City, IA, USA}
\address[c]{Senseable City Lab, Massachusetts Institute of Technology, Cambridge, MA, USA}

\begin{abstract}
{\color{black}
Geospatial algorithms have been developed through \emph{expert-driven discovery}: a researcher spots a limitation, proposes a structural fix, and checks it by hand. With LLM-driven multi-agent systems becoming practical, much of the candidate-generation, evaluation, and selection work can now be largely automated, while the researcher's role shifts upstream toward problem formulation, evaluation design, and interpretation. \nv{We refer to this paradigm as \emph{Self-Evolving Geospatial Algorithms} (SEGA) and propose a framework with three required modules: a \emph{two-loop architecture with embedded geographic knowledge} as the inner-loop engine; an \emph{automatic code formatter} that converts any raw geospatial algorithm into the standardized triple required by the inner loop; and a \emph{simulation-driven training protocol}, in which evolution runs only on synthetic data with known ground truth, and the discovered algorithm is applied unchanged to real data, keeping real-world results out-of-sample.} We use \emph{spatial regime discovery} as a case study, seeded with the Two-stage K-Models algorithm. Evolution uses the Rand Index against the ground-truth partition as fitness signal; the frozen algorithm is then evaluated on 150 simulated grids and 11 real-world datasets across climate, health, hydrology, and politics. \nv{The evolved algorithm outperforms its seed, knowledge-free baselines, and four classical regionalization methods. Ablations further indicate that the channel of geographic-knowledge injection, structured retrieval versus prompt-level hints, is itself load-bearing.} \rw{In this case study, LLM-assisted search sped up the inner loop; whether it does so for other seeds and task families remains open. The geographer's role --- asking the right questions, designing trustworthy evaluation --- grows more central.}
}
\end{abstract}

\end{frontmatter}


\section{Introduction}\label{sec:intro}
Geospatial modeling sits at the core of geospatial research, supporting tasks such as spatial partitioning \citep{anselin2024endogenous,duque2007supervised,assuncao2006efficient}, prediction \citep{brunsdon1998geographically,fotheringham2002geographically,fotheringham2017multiscale}, {\color{black}spatial dependence modeling \citep{anselin1988spatial,lesage2009introduction,anselin1995local},} and uncertainty quantification \citep{lou2025geoconformal}. The development of these models has typically grown out of an expert-driven process that brings together theoretical insights, empirical observations, and domain knowledge{\color{black}, often accumulating over decades of methodological refinement \citep{cressie1993statistics,goodchild2004validity}}. {\color{black}Such development can be characterized as \emph{expert-driven discovery}: a researcher identifies a limitation, hypothesizes a structural improvement (e.g., adding spatial weights, restructuring a clustering objective, or proposing a new estimator), implements it, and validates it against established baselines. Each discovery step is conceptually rich but is also human-paced, costly in time, and naturally limited by the cognitive and institutional resources available to any single research group.}

{\color{black}The rapid development of large language models (LLMs), including ChatGPT \citep{singh2025openai}, Gemini \citep{team2024gemini}, DeepSeek \citep{liu2024deepseek}, and Qwen \citep{yang2025qwen3}, together with the rise of LLM-based multi-agent architectures \citep{park2023generative}, suggests that much of the routine engineering work of algorithm development can now be largely automated. Working in this complementary mode, the researcher delegates the work of generating candidate programs, evaluating them, and keeping the better ones to an LLM-driven system, while retaining responsibility for deciding what to optimize, setting up the evaluation, and making sense of what comes out. Working in this mode, researchers spend less time writing and debugging small algorithmic variants by hand, and more time on the upstream questions: what is a good solution to this problem, what fitness signal captures it, and what do the resulting algorithms tell us. The two modes are best seen as complementary: expert-driven discovery continues to mark out the design space and the objectives that matter, while LLM-assisted search systematically explores parts of that space that would be tedious to cover manually \citep{novikov2025alphaevolve,romera2024mathematical}. Within geospatial science, the convergence of GeoAI \citep{janowicz2020geoai,mai2022opportunities} and LLM-based agents is opening up new directions for spatial analysis and knowledge discovery.}

{
It is useful to distinguish at the outset between two complementary lines of LLM-based research now active in geospatial science. The first treats the LLM as an \emph{orchestrator} of a fixed library of pre-existing geospatial tools and algorithms: the model interprets a user request, selects appropriate operations, and emits glue code that invokes them, while the underlying spatial algorithms remain entirely human-authored. We refer to this as the \emph{usage layer} of the algorithm stack. The second treats the LLM as an \emph{author} that proposes and mutates algorithm source code itself, generating new algorithmic logic rather than choosing among existing tools. We refer to this as the \emph{design layer}. The present paper, together with the GeoEvolve framework it builds on \citep{luo2025geoevolve}, sits firmly at the design layer: the toolbox is the object that gets evolved, not the menu that gets chosen from. Section~\ref{sec:related_works} reviews both lines in detail and identifies the specific inner-loop engines we adopt; here we only flag the distinction so as to keep the technical premise of the rest of the paper visible.
}

{\color{black}
The central message of the present paper is that the combination of these two modes describes a noticeable shift in how geospatial algorithms can be developed, and that this shift deserves to be named and characterized as such. \nv{We refer to this LLM-assisted paradigm as \emph{Self-Evolving Geospatial Algorithms} (SEGA), denoting a structured, evaluation-driven search over the space of executable geospatial programs, in which candidate algorithms are continuously generated, evaluated, and selected by an LLM-powered multi-agent system.} We propose a working hypothesis to guide its scope: \emph{if the optimization objective of a geospatial problem can be specified clearly enough for computational evaluation, the inner loop of algorithm development can be largely delegated to a multi-agent LLM system, while the outer loop of formulating the problem and interpreting the results remains the responsibility of the researcher.} We do not claim that algorithm design has become trivial; rather, we observe that, given a well-defined objective, LLM-driven evolutionary systems can systematically explore parts of the design space that are difficult to traverse manually.
}

{
A natural starting point for realizing SEGA is the recently proposed GeoEvolve framework \citep{luo2025geoevolve}, which extends flat, single-loop, domain-agnostic evolutionary engines such as OpenEvolve \citep{openevolve} by wrapping the inner LLM-driven code-evolution loop with an outer dialogue that retrieves geographic knowledge from a curated corpus and injects it into the next mutation. This two-loop architecture with embedded geographic knowledge is, in our view, the necessary substrate for any practical \red{SEGA} system, and we adopt it as the inner-loop engine of our framework without redesign. Two further extensions, however, are required before this engine can be used routinely on new geospatial problems. The first concerns \emph{the input side}: raw geospatial algorithms are written in heterogeneous styles, and integrating each new seed by hand re-introduces a manual-engineering bottleneck that the inner loop is supposed to remove. An LLM-driven \emph{automatic code formatter} that converts any raw geospatial algorithm into the standardized triple (evolvable program, evaluator module, configuration file) required by the inner loop is therefore needed to make the framework reusable across tasks. The second concerns \emph{the evaluation side}: when evolution is driven directly by the algorithm's performance on the very data on which its output will be reported, real-world metrics become both training target and evaluation target, and the resulting numbers tell us less than they appear to. A \emph{simulation-driven training protocol with a controlled data-generating process}, in which evolution is run only on synthetic data with a known ground truth and the discovered algorithm is then applied unchanged to real data, is therefore needed to keep real-world results as out-of-sample evidence rather than direct optimization targets. The three pieces together, GeoEvolve's two-loop architecture, an automatic code formatter, and a simulation-driven training protocol, constitute the conceptual framework developed in the remainder of this paper.
}

{
To make this paradigm shift concrete rather than abstract, the bulk of the paper develops a case study on \emph{spatial regime discovery}, taking the Two-stage K-Models algorithm \citep{guo2023extending} as the seed algorithm $\mathcal{A}_0$. We chose this case study for three reasons: it represents the current state of the art among non-LLM spatial regime methods, its two-stage structure of initial over-partitioning followed by region merging exposes a small number of clearly evolvable modules that an LLM-driven inner loop can target, and its piecewise-linear regime assumption admits a clean simulated ground truth required by the train-on-simulation / apply-on-real evaluation protocol described below. \rw{As a single case study, it shows that the framework works on this task; applying it to further seeds and task families in future work is needed to establish that it generalizes.} A more complete account of the spatial regime literature and Two-stage K-Models is given in Section~\ref{subsec:spatial_regime_background}.

Within the case study, evolution happens exclusively on simulated grids using a single bounded fitness signal (Rand Index against the ground-truth partition); the resulting algorithm is then applied unchanged to 11 real-world datasets across climate, health, hydrology, and politics, where SSR, SSR Reduction, RMSE, and Variance Ratio are used as evaluation metrics \rv{disjoint from the fitness signal}. On top of this, we conduct a structured ablation that varies both the level and the specificity of geographic-knowledge injection (details in Section~\ref{subsec:variants}), so as to test how (not merely whether) domain knowledge should enter the inner loop of the two-loop architecture.
}

\nv{The remainder of the paper is organized as follows. Section~\ref{sec:related_works} reviews related work on LLM-based autonomous agents, distinguishes earlier LLM-based search over short programs and expressions (Section~\ref{subsec:function_search}) from the more recent AlphaEvolve paradigm of LLM-driven algorithm discovery (Section~\ref{subsec:engines_background}) on which this paper builds, and closes with a background subsection on the spatial regime literature that motivates the case study (Section~\ref{subsec:spatial_regime_background}). Section~\ref{sec:methods} develops the SEGA framework, with a conceptual subsection that frames the paradigm and formalizes it as a quintuple (Section~\ref{sec:concept}) and a detailed subsection that implements the three required modules (Section~\ref{sec:three_modules}). Section~\ref{sec:exp_design} presents the spatial regime case study together with its experimental design, including the problem definition, the simulation experiment with its six knowledge-injection variants, and the real-world experiment with its four classical regionalization baselines. Section~\ref{sec:results} reports the empirical results on both simulated and real-world data. Section~\ref{sec:discussion} discusses what the discovered algorithm modified, the insights and complementarity with expert-driven design, and limitations, framed around the broader question of how autonomous geospatial algorithm discovery should proceed. Section~\ref{sec:conclusion} concludes.}

\section{Related Works}\label{sec:related_works}
\subsection{LLM-based Autonomous Agents}
LLM-based autonomous agents have begun to be explored in geographical information science (GIScience), where agents are designed to support spatial analysis and mapping workflows. \teal{Across this line of work the LLM operates at what we call the \emph{usage layer} of the algorithm stack: it acts as a decision core that interprets user intent and orchestrates a fixed library of pre-existing geospatial tools, while the underlying spatial algorithms remain entirely human-authored.} Early studies use LLMs to translate spatial problems into executable geoprocessing steps, ranging from spatial data retrieval to invoking spatial algorithms \citep{zhang2023geogpt,li2023autonomous}. Multi-agent pipelines have also been used to decompose spatial tasks into planner-worker structures for operations such as shapefile processing and spatial analysis \citep{lin2025shapefilegpt}. Other work investigates automated geospatial data discovery agents that generate and run programs to locate appropriate spatial datasets \citep{ning2025autonomous}. By emulating cartographic workflows, LLM agents can also be guided to design, add, or modify map elements as part of an interactive mapping process \citep{zhang2024mapgpt}. \teal{Related efforts include LLM-powered QGIS spatial-analysis copilots \citep{akinboyewa2025gis}, harness-engineered research agents that span the broader GIScience research lifecycle \citep{zhou2026nora}, and the wider autonomous-GIS research agenda \citep{li2025giscience}.} While these efforts demonstrate the value of LLM agents for assembling and orchestrating existing spatial operations, they typically rely on pre-defined tools rather than directly proposing new algorithmic structures. Recent work suggests that LLMs may also support more open-ended forms of geospatial scientific exploration, in which agents help propose hypotheses, identify candidate spatial patterns, and combine heterogeneous datasets to generate further insight.

\subsection{\nv{LLM-Based Search over Short Programs and Expressions}}\label{subsec:function_search}
\nv{The use of LLMs has prompted reconsideration of how new algorithms and mathematical expressions can be discovered. In contrast to the usage-layer agents reviewed above, the systems considered here operate at the \emph{design layer} of the algorithm stack: the LLM is not selecting among existing algorithms but authoring them, generating and mutating the source code that implements the algorithmic logic itself.}

\nv{An early line of work treats the LLM as a high-level variation operator embedded inside an evolutionary or search loop, applied to compact single-file programs. Eureka \citep{ma2023eureka} used an LLM to evolve dense reward functions for reinforcement learning by editing short candidate programs. LMX \citep{meyerson2024language} explored LLM-driven variation over linguistic genotypes as a generic operator. FunSearch \citep{romera2024mathematical} applied the same idea to mathematical reasoning, using the LLM to mutate short Python functions and reporting novel constructions for problems such as the cap-set problem and online bin-packing. LLM-SR \citep{shojaee2024llm} adapted symbolic regression in the same spirit, treating the discovery of compact mathematical expressions as program search.}

\nv{Common to these systems is that the unit being evolved is a single short function or expression rather than the broader algorithmic framework in which it is embedded: the LLM proposes local edits inside a few-line, single-file program. While these results demonstrate that LLM-driven search can produce novel solutions on well-scoped function-level problems, the mutation envelope is too narrow to express the structural moves required to evolve a full geospatial algorithm with multiple stages, parameters, and data pipelines, for example swapping an initialization strategy, restructuring a clustering objective, or adding a per-region penalty term. The next subsection introduces the more recent line of work that targets this larger unit of evolution.}

\subsection{\nv{LLM-Driven Algorithm Discovery: The AlphaEvolve Paradigm}}\label{subsec:engines_background}
{
\nv{A second, more recent line of work targets the discovery of \emph{full algorithms} rather than short functions, by allowing the LLM to mutate program text spanning multiple files and many lines of logic. We refer to this as the \emph{AlphaEvolve paradigm}. The three systems described below, AlphaEvolve as the original proposal, OpenEvolve as its open-source implementation, and GeoEvolve as its first geospatial specialization, jointly characterize this paradigm. Because this paper uses OpenEvolve and GeoEvolve directly as the inner-loop engine $\Phi$ of Section~\ref{sec:methods} rather than proposing a new engine, this subsection describes what each system actually provides before the case study begins.}

\paragraph{AlphaEvolve} \nv{AlphaEvolve \citep{novikov2025alphaevolve} extends LLM-driven evolution from single short functions to entire multi-file codebases, reporting measurable improvements on long-standing algorithmic problems including matrix multiplication and other open mathematical questions. The methodological move is significant: instead of evolving a short single function of a few dozen lines, the system mutates and recomposes algorithmic structures of hundreds of lines, including initialization strategies, iteration schedules, and helper subroutines. This brings algorithm discovery, rather than merely function discovery, within reach of LLM-driven search.}

\paragraph{OpenEvolve} OpenEvolve \citep{openevolve} is an open-source implementation in the spirit of AlphaEvolve, packaging the same inner loop as a domain-agnostic engine. Given a seed program with clearly marked evolvable blocks, a fitness function, and an iteration budget, OpenEvolve executes a flat loop of \emph{propose mutation $\to$ run $\to$ evaluate $\to$ select}. It is intentionally domain-agnostic: it carries no notion of geographic space, no curated scientific corpus, and no mechanism for steering the search beyond whatever text happens to be in the prompt. Within this paper, OpenEvolve serves both as a standalone baseline, used directly with or without hand-written geographic-knowledge hints in the prompt, and as the underlying mutation engine that GeoEvolve wraps.

\paragraph{GeoEvolve} GeoEvolve \citep{luo2025geoevolve} is the first geospatial specialization of the AlphaEvolve paradigm: it wraps the OpenEvolve loop in three agent-level modules so that geographic domain knowledge can enter each round of evolution through structured retrieval rather than only through static prompts. A \emph{Code Analyzer} is a diagnostic LLM agent that reads the current evolved code together with its evaluation metrics and outputs (in strict JSON) which pieces of geographic knowledge appear missing or misapplied, along with search queries for the knowledge retriever and flags indicating whether code examples or theoretical background are needed. A \emph{Knowledge Retriever (GeoKnowRAG)} maintains a curated corpus of geospatial documents and returns ranked passages relevant to the Analyzer's queries through retrieval-augmented generation. The corpus, embedding model, vector store, and retrieval procedure are all configurable; the specific choices used in this paper are documented in \ref{appendix:geoevolve_config}. Finally, a \emph{Geo-informed Prompt Generator} assembles the retrieved knowledge passages, the analyzer's diagnostic notes, and the current code into a structured prompt that is fed back into the next OpenEvolve mutation step. The result is a two-loop architecture in which geographic knowledge enters the search at the retrieval level, conditioned on the current state of the evolved code, rather than only as fixed prompt context.

\paragraph{Implication for this paper} The distinction between OpenEvolve (prompt-level knowledge only, if any) and GeoEvolve (prompt-level $+$ retrieval-level geographic knowledge) defines a natural design space for ablation. Section~\ref{subsec:variants} uses this design space to construct six algorithm variants spanning no knowledge, a general prompt hint, a task-specific prompt hint, multi-agent dialogue without retrieval, and multi-agent dialogue with full GeoKnowRAG, in addition to the unevolved seed. We use these six variants across the case study to test the hypothesis that the \emph{channel} of knowledge injection (prompt versus retrieval) is itself a load-bearing design choice.
}

\subsection{\teal{Spatial Regime Discovery}}\label{subsec:spatial_regime_background}
{
Because the case study (Section~\ref{sec:exp_design}) is grounded in spatial regime discovery, we briefly review the relevant methodological background here. Spatial regime modeling seeks to partition a study area into contiguous subregions within which the data-generating process relating covariates to a response variable is locally homogeneous but differs across regions \citep{anselin1988spatial,anselin2024endogenous}. It is a central methodological tool for handling spatial heterogeneity \citep{fotheringham2017multiscale} and sits at the intersection of three long-standing traditions: contiguity-preserving regionalization \citep{openshaw1977geographical,openshaw1995algorithms,assuncao2006efficient,duque2007supervised}, geographically weighted regression \citep{brunsdon1998geographically,fotheringham2002geographically,fotheringham2017multiscale}, and graph-based spatial clustering followed by regression \citep{helbich2013data}. Recent work has unified these threads by embedding regression objectives directly into the partitioning procedure: \citet{anselin2024endogenous} extend SKATER with a regression-SSR cut criterion (Skater-reg), while \citet{guo2023extending} propose Two-stage K-Models, Regional-K-Models, and a regression-aware variant of the Automatic Zoning Procedure. Among these, the Two-stage K-Models algorithm is reported by its authors to ``largely outperform existing approaches on model fitting, region reconstruction and coefficient estimation'' \citep{guo2023extending}, making it the strongest non-LLM baseline currently available for spatial regime discovery. This is the algorithm we adopt as the seed $\mathcal{A}_0$ throughout the case study; the rationale for this choice was previewed in the Introduction.
}

\section{\nv{Methods: The SEGA Framework}}\label{sec:methods}

\nv{This section develops the SEGA framework introduced in Section~\ref{sec:intro}. Section~\ref{sec:concept} sets up the concept and the division of labor between researcher and agent, formalized as a quintuple. Section~\ref{sec:three_modules} then specifies each of the three required modules in turn.}

\subsection{\nv{Conceptualizing SEGA: From Paradigm Shift to Quintuple}}\label{sec:concept}

The basic idea of SEGA is to treat algorithm development not as a replacement for expert-driven design but as a complement to it: alongside the conceptual moves researchers make by hand, parts of algorithm refinement can be carried out as a structured, evaluation-driven search over executable programs. Two observations motivate this view. First, the historical record of geospatial algorithm development is dominated by what we call \emph{expert-driven discovery}: discrete, conceptually rich modifications that come out of a researcher noticing a limitation and proposing a fix (for example, adding spatial weights into a regression model, replacing a Euclidean dissimilarity with a contiguity-aware one, or adding a penalty against spatially fragmented partitions). These discoveries are valuable but slow, and the design space between them is large. Second, LLMs and LLM-based multi-agent systems now provide the technical means to automate large parts of the candidate-generation, evaluation, and selection process. The resulting division of labor is straightforward: the agent system handles the generation, evaluation, and selection of variants on top of any given seed algorithm, while the researcher retains responsibility for formulating the problem, designing the objective, defining the evaluation, and interpreting the result. SEGA, in this sense, is an attempt to make use of the second observation without giving up the first.

\begin{figure}[H]
 \centering
 \includegraphics[width=\linewidth]{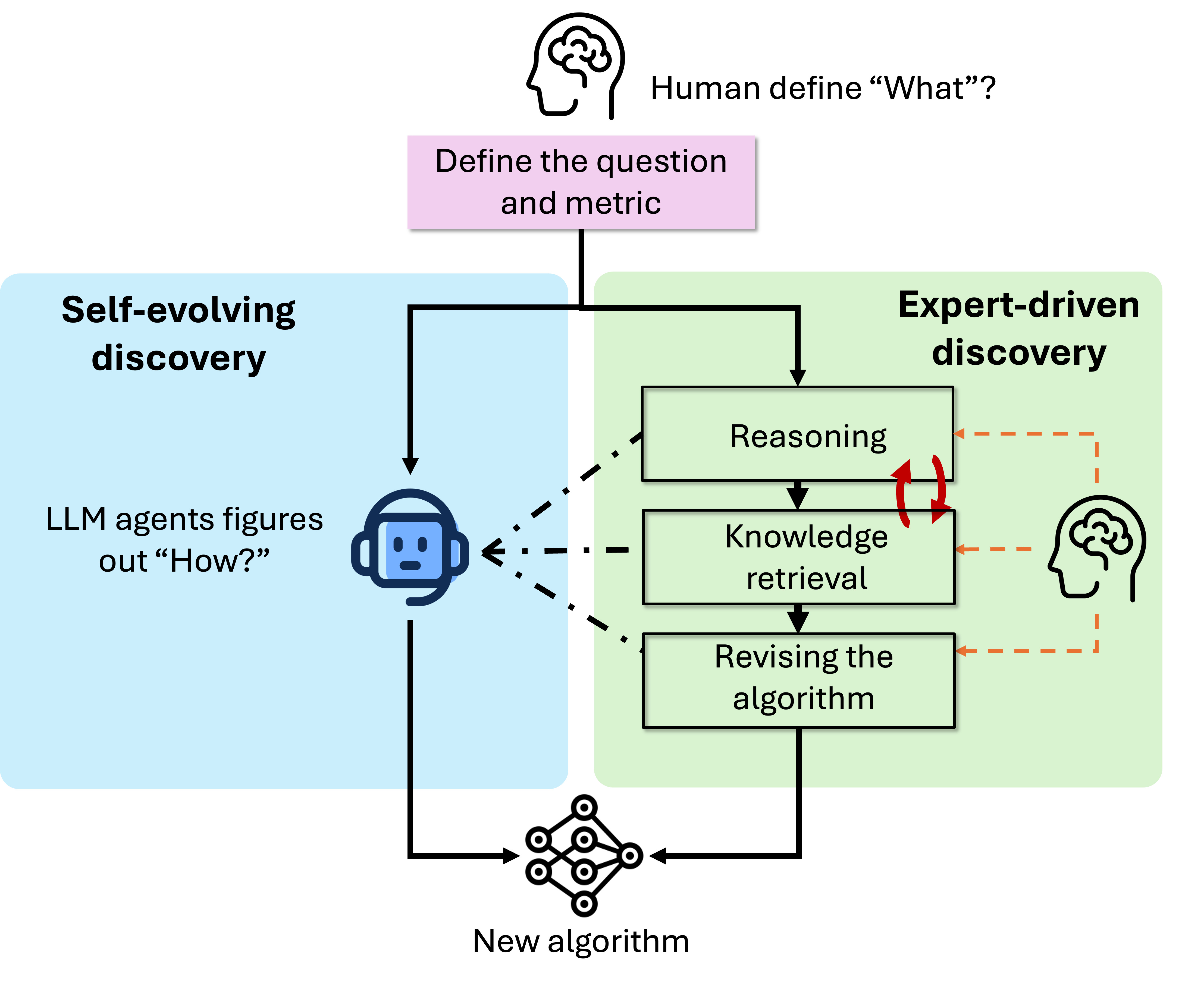}
 \caption{A schematic comparison of expert-driven discovery (green panel) and self-evolving discovery (blue panel). Expert-driven discovery typically advances through conceptual innovations contributed by individual researchers and groups, whereas self-evolving discovery follows a more centralized, metric-driven evolutionary trajectory. In SEGA, the inner loop of \emph{how} to refine an algorithm is largely delegated to the agent, while the outer loop of \emph{what} question to ask remains with the researcher. The two paradigms are complementary rather than mutually exclusive.}
 \label{fig:self-evolving-comparison}
\end{figure}

Figure~\ref{fig:self-evolving-comparison} contrasts the two paradigms in stylized form. In the expert-driven paradigm (right panel), progress relies on conceptual innovations from researchers; each contribution typically requires identifying a limitation, hypothesizing a structural modification, implementing and testing it, and iterating. This process is inherently human-paced and is naturally shaped by available time and disciplinary context. In the self-evolving paradigm (left panel), candidate algorithms are continuously refined through automated mutation, evaluation, and selection within a unified computational framework, but the researcher still owns the upstream decisions: what to optimize, what counts as a good answer, and what the resulting algorithms tell us. Both modes have value: the former provides conceptual depth, the latter offers breadth of exploration.

\paragraph{Formalization as a quintuple} We formalize SEGA as a quintuple $(\mathcal{A}_0, \mathcal{D}, \mathcal{E}, \mathcal{K}, \Phi)$. Here $\mathcal{A}_0$ is the \textbf{seed algorithm}, an initial executable program implementing a baseline geospatial method; $\mathcal{D} = \{(s_i, \mathbf{X}_i, Y_i)\}_{i=1}^N$ is the \textbf{spatial dataset}, with $s_i \in \mathbb{R}^2$ the spatial coordinates, $\mathbf{X}_i \in \mathbb{R}^K$ the explanatory variables, and $Y_i$ the response variable; $\mathcal{E}: \mathcal{A} \times \mathcal{D} \rightarrow \mathbb{R}^m$ is the \textbf{evaluation protocol}, a mapping from an algorithm--dataset pair to an $m$-dimensional performance vector that provides a fitness signal for evolutionary selection; $\mathcal{K}$ is the \textbf{geographic knowledge base}, a curated corpus of spatial theories, statistical methods, and established geographic principles; and $\Phi$ is the \textbf{evolution engine}, an LLM-powered operator that proposes structural variants of candidate algorithms guided by performance feedback and domain knowledge. After $T$ rounds the system returns the discovered algorithm $\mathcal{A}^\star$, its performance profile $\mathcal{E}(\mathcal{A}^\star, \mathcal{D})$, and (if logging is enabled) the evolution trace.

\begin{figure}[H]
 \centering
 \includegraphics[width=\linewidth]{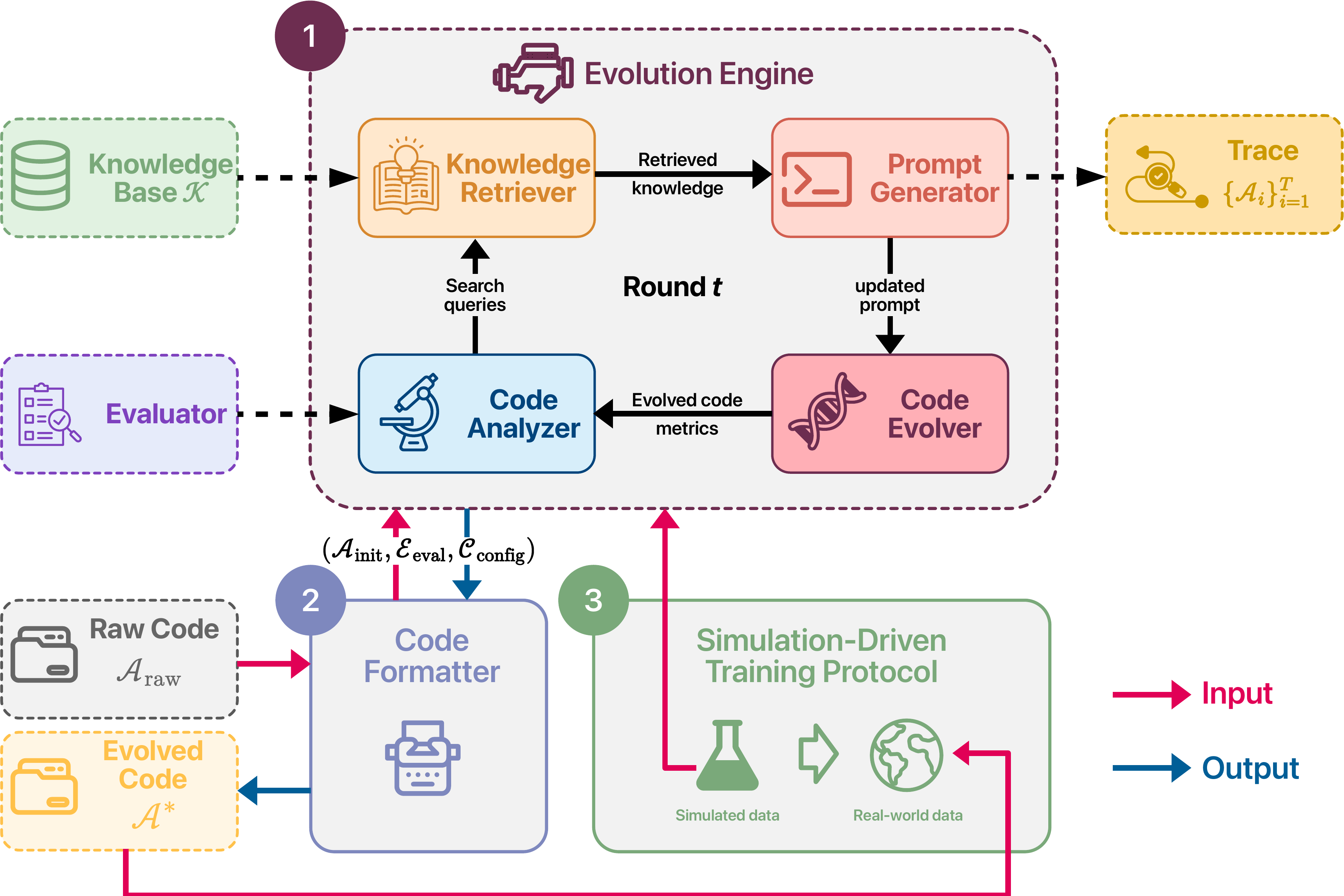}
 \caption{Schematic of the SEGA framework, organized around the three required modules marked by the \modOne, \modTwo, and \modThree badges. \modOne~the \emph{Evolution Engine} $\Phi$ (top dashed panel): a two-loop architecture in which four LLM agents collaborate within each round $t$, the Code Analyzer reads the current evolved code and its evaluation metrics; the Knowledge Retriever queries the curated Knowledge Base $\mathcal{K}$ with the analyzer's search queries and returns retrieved knowledge; the Prompt Generator assembles that knowledge into the next mutation prompt; the Code Evolver produces the next candidate program. The Evaluator and Knowledge Base sit outside the engine; the engine returns the evolution trace $\{\mathcal{A}_i\}_{i=1}^T$. \modTwo~the \emph{Code Formatter} (bottom-left panel): converts user-supplied raw code $\mathcal{A}_{\text{raw}}$ into the standardized triple $(\mathcal{A}_{\text{init}}, \mathcal{E}_{\text{eval}}, \mathcal{C}_{\text{config}})$ consumed by the engine, and rewrites the discovered algorithm $\mathcal{A}^\star$ back to the user-facing format. \modThree~the \emph{Simulation-Driven Training Protocol} (bottom-right panel): the engine sees only simulated data during evolution; real-world data is used only post-evolution as held-out evaluation. Pink arrows denote inputs, blue arrows denote outputs.}
 \label{fig:paradigm_overview}
\end{figure}

\begin{figure}[H]
 \centering
 \includegraphics[width=\linewidth]{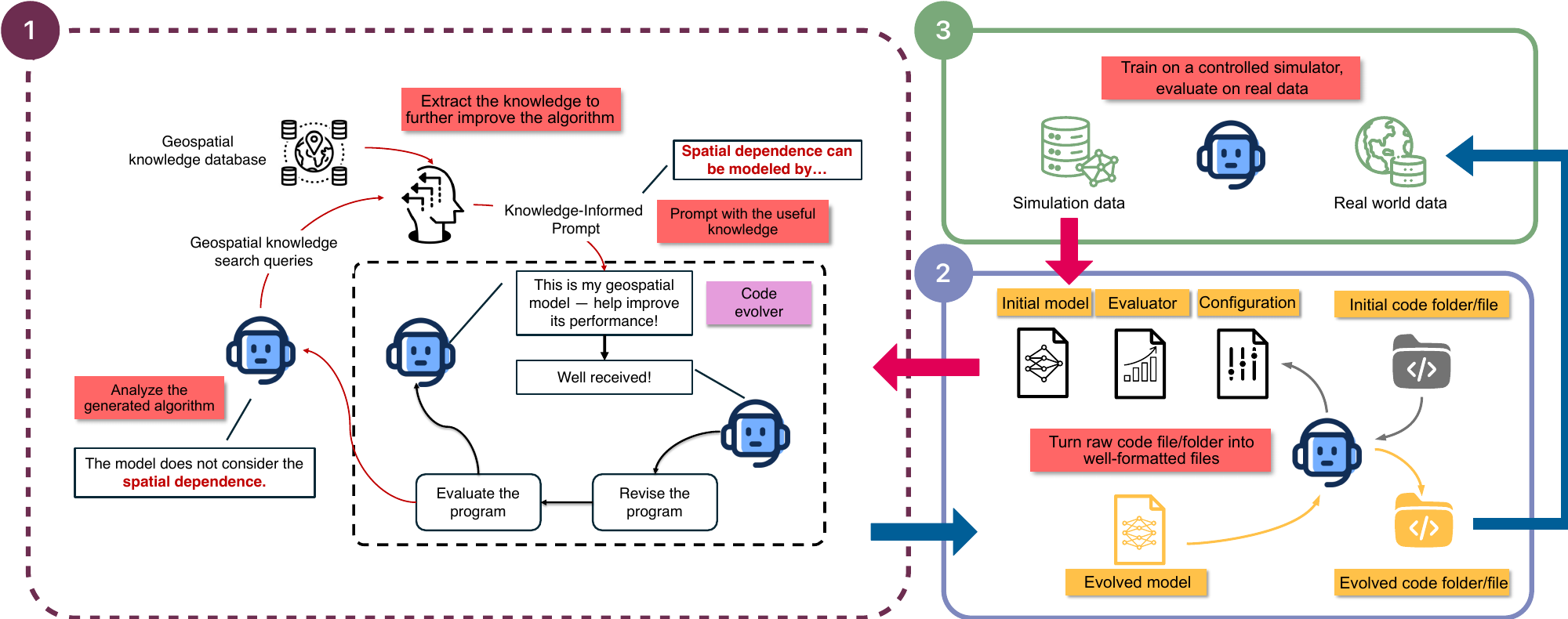}
 \caption{Agent-level details of each SEGA module, complementing the abstract block diagram of Figure~\ref{fig:paradigm_overview}. \modOne~Inside the \emph{Evolution Engine} (left, purple dashed panel): the four agents collaborate in one round of evolution, the Code Evolver submits the current geospatial model (``This is my geospatial model, help improve its performance!''); the Code Analyzer inspects the produced program, diagnoses a missing geographic consideration (``The model does not consider the spatial dependence''), and emits search queries to the geospatial knowledge database; the Knowledge Retriever returns relevant passages (e.g., ``Spatial dependence can be modeled by\dots''); the Prompt Generator wraps the retrieved knowledge into a knowledge-informed prompt; the Code Evolver revises the program, and the inner loop ``Revise the program $\to$ Evaluate the program'' iterates. \modTwo~Inside the \emph{Code Formatter} (bottom-right, blue panel): an LLM-driven formatting agent turns a raw code folder/file into the standardized triple of Initial model, Evaluator, and Configuration (together with the corresponding code folder/file), and re-emits the discovered algorithm as a user-facing evolved model and code folder. \modThree~Inside the \emph{Simulation-Driven Training Protocol} (top-right, green panel): the algorithm being evolved sees only simulated data during training and is then applied unchanged to real-world data for held-out evaluation. Pink arrows denote inputs to the engine, blue arrows denote outputs.}
 \label{fig:agent_dialogue_example}
\end{figure}

\nv{\paragraph{Humans define `what', agents figure out `how'} The quintuple makes explicit which components remain a researcher's responsibility and which can be delegated to the agent, and the division it draws is exactly the one shown in Figure~\ref{fig:self-evolving-comparison}: the researcher defines the \emph{what}, the question and the metric that counts as a good answer, while the agent (in the self-evolving mode) figures out the \emph{how}. Concretely, the researcher chooses $\mathcal{A}_0$ (which existing algorithm to use as the starting point), constructs $\mathcal{D}$ (which simulated and real-world data to use), designs $\mathcal{E}$ (what counts as a good solution and how to measure it), and curates $\mathcal{K}$ (what geographic knowledge the system can draw on); the agent, via $\Phi$, handles the inner loop, generating candidate variants of $\mathcal{A}_0$, running them on $\mathcal{D}$, scoring them with $\mathcal{E}$, and selecting which to keep. Three modules are required to turn this abstract quintuple into a system that can actually be applied to a new geospatial problem, and they are exactly the three pieces flagged in the Introduction: the Evolution Engine \modOne~(Section~\ref{subsec:module1}), the Code Formatter \modTwo~(Section~\ref{subsec:module2}), and the Simulation-Driven Training Protocol \modThree~(Section~\ref{subsec:module3}). Figure~\ref{fig:paradigm_overview} provides the abstract block-level architecture of SEGA; Figure~\ref{fig:agent_dialogue_example} expands each of the three modules to show, in an infographic form, the agent-level workings inside them.}

\subsection{\nv{Implementing SEGA: Three Required Modules}}\label{sec:three_modules}

\subsubsection{\nv{Module \modOne: Two-Loop Architecture with Embedded Geographic Knowledge}}\label{subsec:module1}

The first module specifies how the evolution engine $\Phi$ is organized so that geographic domain knowledge $\mathcal{K}$ can be brought to bear on each round of mutation. We require a two-loop architecture: an \emph{inner loop} that performs LLM-driven code evolution (propose mutation $\to$ run $\to$ evaluate $\to$ select), wrapped by an \emph{outer loop} that analyzes the current evolved code, retrieves relevant geographic knowledge from $\mathcal{K}$, and feeds the retrieved knowledge back into the next inner-loop mutation as a structured prompt. The inner loop is responsible for exploring the space of algorithmic variants; the outer loop is responsible for steering that exploration with established geographic theory. Formally, the inner loop computes
\begin{equation}
 \mathcal{A}^\star = \arg\min_{\mathcal{A}\in \Phi^T(\mathcal{P}_0)} \mathcal{L}(\mathcal{A}, \mathcal{D}_\text{sim}),
\end{equation}
where $\mathcal{P}_0$ is the initial candidate set seeded from $\mathcal{A}_0$, $\Phi^T(\cdot)$ is the evolutionary process over $T$ iterations, $\mathcal{L}$ is the fitness function, and $\mathcal{D}_\text{sim}$ is the simulated dataset used during evolution (the use of $\mathcal{D}_\text{sim}$ rather than $\mathcal{D}$ is enforced by Module~\modThree below).

\paragraph{Adopted realization: GeoEvolve} The two-loop architecture is the central methodological innovation of GeoEvolve \citep{luo2025geoevolve}, distinguishing it from flat, single-loop, domain-agnostic engines such as OpenEvolve \citep{openevolve} in which geographic knowledge can enter only as static prompt text. We adopt GeoEvolve as our off-the-shelf inner-loop engine, with its default agent architecture, prompts, and hyperparameters from the original paper (background on the four agents is given in Section~\ref{subsec:engines_background}); the top dashed panel (\modOne) of Figure~\ref{fig:paradigm_overview} shows how the four agents are wired together within the Evolution Engine $\Phi$. \nv{In practice we use the open-source Python package \texttt{geoevolve} released alongside \citet{luo2025geoevolve}, which packages the entire two-loop architecture (Code Evolver, Code Analyzer, Knowledge Retriever, and Geo-informed Prompt Generator) behind a single configuration interface, so that adopting GeoEvolve as the inner-loop engine requires no re-implementation.}

\paragraph{Design space of knowledge integration} A practical question remains open even after adopting GeoEvolve: \emph{how} geographic knowledge enters $\Phi$ through the outer loop. We characterize the design space along two axes. The first axis is the \emph{level} at which knowledge is injected: prompt-level injection encodes domain knowledge as natural-language instructions in a static prompt, whereas retrieval-level injection dynamically retrieves relevant knowledge from a curated corpus and conditions it on the current state of the evolved code. The second axis is the \emph{specificity} of the injected knowledge: none, general (e.g., Tobler's first law, spatial autocorrelation, scale effects), or task-specific (e.g., contiguity penalties, region-size balancing for spatial regimes). The combination of these two axes defines the six algorithm variants used in our case-study ablation (Section~\ref{subsec:variants}).

\subsubsection{\nv{Module \modTwo: Automatic Code Formatter}}\label{subsec:module2}

The second module operates on the seed algorithm $\mathcal{A}_0$: it specifies how a raw geospatial algorithm, written in whatever style its authors prefer, is turned into the standardized input that the inner-loop engine $\Phi$ requires. Without this step every new seed requires manual engineering before evolution can begin; with it, \rw{the same evolutionary infrastructure is designed to be reusable across different seeds and tasks with little or no manual adaptation. We state this as a property of the design rather than a demonstrated one: the formatter is exercised here on a single seed (Section~\ref{sec:exp_design}), and whether it generalizes across algorithm families remains untested (Section~\ref{subsec:limitations}).}

We define the formatting problem as a transformation that takes a raw geospatial algorithm $\mathcal{A}_{\text{raw}}$ as input and returns a standardized triple
\begin{equation}
 \text{Format}(\mathcal{A}_{\text{raw}}) \rightarrow (\mathcal{A}_{\text{init}}, \mathcal{E}_{\text{eval}}, \mathcal{C}_{\text{config}}),
\end{equation}
in which $\mathcal{A}_{\text{init}}$ is an \textbf{initial program} with clearly marked evolvable blocks and a standardized entry function that returns predictions together with any auxiliary outputs (region labels, coefficients, etc.) needed by the evaluation step; $\mathcal{E}_{\text{eval}}$ is an \textbf{evaluator module} that implements the evaluation protocol $\mathcal{E}$ and computes the fitness signal and the held-out evaluation metrics from the algorithm's outputs; and $\mathcal{C}_{\text{config}}$ is a \textbf{configuration specification} that records the evolutionary hyperparameters (population size, number of iterations, mutation strategy) and any task-specific settings (target number of regions, convergence criteria, soft constraints). This standardized triple acts as a contract between the candidate algorithm and the inner-loop engine: regardless of how the underlying algorithm is written, each candidate exposes a uniform interface for execution, evaluation, and comparison.

In our implementation, the formatting step itself is performed automatically by an LLM-driven parser: given a raw script that solves a geospatial task in some style, the parser identifies the core algorithmic logic, marks the evolvable blocks, generates the standardized entry function, writes the evaluator module against the user-specified metrics, and emits a configuration file together with metadata for traceability. Researchers therefore only need to provide a working implementation of their geospatial algorithm; the engineering work of preparing it for evolutionary search is handled by the formatter. We use this formatter to prepare the Two-stage K-Models seed used throughout the case study. A worked example of the formatter operating on the Two-stage K-Models seed is provided in \ref{appendix:formatter_example}, showing the raw input, the resulting evolvable program $\mathcal{A}_{\text{init}}$ with marked blocks, the generated evaluator module $\mathcal{E}_{\text{eval}}$, and the configuration $\mathcal{C}_{\text{config}}$.

\subsubsection{\nv{Module \modThree: Simulation-Driven Training Protocol with a Controlled DGP}}\label{subsec:module3}

The third module operates on the pair $(\mathcal{D}, \mathcal{E})$: it specifies what dataset the inner loop is allowed to see during evolution, and what metric the inner loop is asked to optimize. We argue that these two choices should be made together, and at the framework level, because they determine whether the resulting algorithm is being evolved against something geographically meaningful or simply against statistical convenience. Concretely, we propose three principles, all centered on the idea of \emph{training on a controlled simulator and evaluating on real data}.

\paragraph{Principle 1: Run the inner loop only on synthetic data} The inner loop $\Phi$ is run only on a synthetic dataset $\mathcal{D}_\text{sim}$ whose ground-truth structure $\mathcal{S}^\star$ is known by construction (for example, a known partition, a known surface, or a known set of intervals). $\mathcal{S}^\star$ gives a clear, unambiguous fitness signal, and the real-world data is never seen by the mutation, selection, or knowledge-retrieval steps, so real-world results count as out-of-sample evidence rather than as a target the system has quietly been optimizing against.

\paragraph{Principle 2: Apply the discovered algorithm directly, without further tuning on real data} After $T$ rounds, the discovered algorithm $\mathcal{A}^\star$ is frozen and applied unchanged to each real-world dataset. No hyperparameters are retuned, no additional evolution rounds are run, and no knowledge retrieval is done on the real-world side. Any performance difference observed on real-world data can then be attributed to the structure of $\mathcal{A}^\star$ itself, rather than to per-dataset adjustments.

\paragraph{Principle 3: Use different metrics for driving the search and for reporting results} The metrics in the system play two distinct roles. A \emph{fitness signal} drives the inner loop $\Phi$; its purpose is to steer the search toward better algorithms. \emph{Evaluation metrics} describe the discovered algorithm after the search is over; their purpose is to tell the reader how well it actually performs. We recommend keeping these two sets separate, because using the same metric in both roles is uninformative: any algorithm produced by an evolutionary search that explicitly optimizes a metric $M$ will, by construction, score well on $M$.

\paragraph{What makes a good fitness signal?} Within these three principles, a useful fitness signal is a \emph{bounded, dimensionless} metric that compares the algorithm's output to the ground-truth structure $\mathcal{S}^\star$ directly. Examples include clustering-agreement indices such as the Rand Index for partitioning tasks, coverage--width scores for prediction-interval tasks, and bounded ranking measures for retrieval tasks. Such fitness signals are bounded (the search has a stable target), robust to trivial structural choices (they cannot be gamed by simply producing more, smaller regions, as SSR-family metrics can), and on a different scale from the usual evaluation metrics, which keeps the evaluation set genuinely held-out.

One \rw{intended} consequence of these principles is that the same protocol \rw{should be} reusable, with minor adjustments, across very different SEGA tasks: only $\mathcal{D}_\text{sim}$, $\mathcal{S}^\star$, the fitness signal, and the evaluation metrics need to be re-specified per task, while the surrounding inner-loop machinery is held fixed. \rw{This too is a consequence of the design rather than a result: the protocol is instantiated here for one task (Section~\ref{subsec:sim_design}), and whether the re-specification is as contained as the argument suggests is a question for the further demonstrations called for in Section~\ref{subsec:limitations}.}

\section{\nv{Case Study: Spatial Regime Algorithm Discovery}}\label{sec:exp_design}

\nv{In this section we apply SEGA to spatial regime discovery. Section~\ref{subsec:problem_def} states the problem and the seed algorithm. Section~\ref{subsec:sim_exp} describes the simulation experiment used to evolve the algorithm, including the simulator and the fitness signal. Section~\ref{subsec:real_exp} describes the real-world experiment used as held-out evaluation, including the eleven datasets across four domains. Section~\ref{subsec:comparison_methods} then specifies the ten algorithms compared throughout both experiments: six Two-stage K-Models variants (the unevolved seed and five evolved variants spanning a $2{\times}3$ knowledge-injection design) and four classical regionalization baselines. The same ten algorithms are evaluated under the same protocol in both the simulation experiment (Section~\ref{subsec:sim_results}) and the real-world experiment (Section~\ref{subsec:real_results}).}

\subsection{\nv{Problem Definition and Seed Algorithm}}\label{subsec:problem_def}

\paragraph{Spatial regime modeling}
Let $\mathcal{S}\in R^2$ be a geographical area partitioned into $R$ mutually exclusive and exhaustive subregions:
\begin{equation}
 \mathcal{S}=\bigcup_{r=1}^R\mathcal{S}_r, \mathcal{S}_r\cap\mathcal{S}_{r'}=\emptyset\text{ for } r\neq r'.
\end{equation}
For an observational unit $i$ located at $s_i\in\mathcal{S}$, a spatial regime exists when the data generation process differs across these spatial partitions:
\begin{equation}
 Y_i = \mathcal{F}_{r(i)}(X_i)+\epsilon_i,
\end{equation}
where $\mathcal{F}_r: \mathbb{R}^K\rightarrow \mathbb{R}$ is a regime-specific functional relationship, $\epsilon_i$ is an error term, $Y_i$ is the dependent variable, and $X_i=\{x_{i,1},...,x_{i,K}\}$ is the vector of $K$ independent variables for spatial unit $i$. The objective of spatial regime modeling is to reduce within-region structural variability while increasing between-region structural differentiation in the data generation mechanism. \rv{Spatial regime discovery is thus a \emph{descriptive partitioning} task, in the same family as clustering and regionalization: given the observed units of a study area, the algorithm assigns each unit to a regime and fits regime-specific relationships, and the recovered regime structure is itself the analytical result --- a characterization of how the data-generating process varies across the study area. Unlike predictive spatial models, a regime model makes no prediction at unobserved locations and assigns no regime to units outside the data it partitions; its outputs are evaluated by how well the partition accounts for structural heterogeneity within the given data, in the same way that a fitted regression is summarized by its goodness of fit. This distinction determines how the algorithms in this paper are evaluated (Section~\ref{subsec:real_design}).}

\paragraph{Seed algorithm: Two-stage K-Models}
We use the Two-stage K-Models (2KModels) algorithm \citep{guo2023extending} as the seed algorithm $\mathcal{A}_0$. 2KModels is a two-stage method that first partitions a region into $M$ subregions $\{\mathcal{S}_1,...,\mathcal{S}_M\}$ (with $M>R$ a hyperparameter controlling the number of initial clusters), and then iteratively merges these subregions into geographically connected regions until $R$ regions are reached. As discussed in Section~\ref{subsec:spatial_regime_background}, 2KModels is the current state of the art among non-LLM spatial regime methods.

\subsection{\nv{Simulation Experiment}}\label{subsec:sim_exp}

\subsubsection{\nv{Design}}\label{subsec:sim_design}

We instantiate the train-on-simulation / apply-on-real protocol of Section~\ref{subsec:module3} for spatial regime discovery as follows.

\paragraph{Simulated datasets}
We generated a series of datasets to evaluate the performance of both the original and evolved versions of the spatial regime algorithm under known ground-truth spatial regimes. Given the response variable $Y$ and explanatory variables $X_1$, $X_2$, the data generation process (DGP) is defined as:
\begin{gather}
 Y = \beta_0(u,v) + \beta_1(u,v)X_1 + \beta_2(u,v)X_2 + \epsilon, \\
 \text{where } X_1,X_2\sim \mathcal{U}[0,1), i.i.d,\\
 \epsilon \sim \mathcal{N}(0,\sigma^2), i.i.d.
\end{gather}
where $(u,v)$ is one of 625 georeferenced observations on a 25 $\times$ 25 regular grid, $\mathcal{U}[0,1)$ is the uniform distribution on $[0,1)$, $\mathcal{N}(0,\sigma^2)$ is the normal distribution with mean 0 and variance $\sigma^2$, and $\beta_i(u,v)$ are spatially varying coefficients that change across regions. The grid is divided into five spatially connected regions $\mathcal{S}^\star=\{\mathcal{S}_1^\star,...,\mathcal{S}_5^\star\}$ that constitute the ground-truth partition. The intercept term $\beta_0$ is set to 0, and the covariate coefficients $\beta_1$, $\beta_2$ take values in $\{-2, -1, 0, 1, 2\}$ depending on the subregion of each cell. Three latent spatial regime patterns are used. In the \textbf{Rectangular} pattern, the grid is split into five rectangular regions, each sized $5 \times 25$ units. In the \textbf{Voronoi} pattern, five cells are randomly chosen as seeds and the resulting Voronoi diagram defines five spatially contiguous regions. In the \textbf{Arbitrary} pattern, five cells are randomly chosen as seeds and each seed expands at random to generate five connected regions. We produce 150 grids in total at the high-noise level, evenly split across the three patterns.

\paragraph{Fitness signal: Rand Index}
Following Principle 3 of Module~\modThree (Section~\ref{subsec:module3}), the inner loop $\Phi$ optimizes a single bounded fitness signal that compares the algorithm's output against the ground-truth partition $\mathcal{S}^\star$. We use the Rand Index (RandI):
\begin{equation}
 \text{RandI}=\frac{TP+TN}{TP+FN+FP+TN}
\end{equation}
where TP, FN, FP, and TN denote true positive, false negative, false positive, and true negative pairs of cells, respectively. RandI is bounded in $[0,1]$, robust to inflated $k$ (it penalizes both over- and under-segmentation), and disjoint in scale from the SSR-family metrics \rv{reserved for post-hoc evaluation}. It is the only metric the inner loop $\Phi$ sees: the LLM-driven mutation, selection, and knowledge-retrieval steps do not access SSR, RMSE, or any other downstream metric. \nv{For computational tractability, the inner loop does not evaluate every candidate program on all 150 grids; instead, evolution operates on a fixed subset of three representative grids, one per spatial pattern (Rectangular \#5, Voronoi \#75, Arbitrary \#149), and the per-round fitness reported during evolution is the mean RandI across these three grids. The discovered algorithm $\mathcal{A}^\star$ is then evaluated, frozen, on the full 150-grid pool in Section~\ref{subsec:sim_results}; \rv{the per-round fitness during evolution (mean RandI over the three training grids) and the per-pattern metrics in Table~\ref{tab:baseline_simulation} (means over the full 150-grid pool, which contains those three grids) are therefore aggregated over different grid sets and are not directly comparable; excluding the three training grids from the 150-grid evaluation changes the reported means by less than $0.2$ SSR and $0.001$ ARI.} \rv{\paragraph{Evaluation-harness isolation} The harness prevents a generated candidate from ever seeing the ground-truth partition. A candidate receives only $(X, Y, w, p, \text{min\_size})$ through a fixed entry point and returns a label vector; RandI is computed by the harness outside the candidate's process, and a static scan rejects, with fitness zero, any program that references ground-truth labels, computes label-agreement metrics internally, or performs file or network I/O. Every program generated in the reported runs was additionally audited for ground-truth references; none were found. The discovered algorithm's own internal objective is likewise label-free --- it selects among candidate partitions by post-merge SSR under a connectivity constraint --- so the frozen algorithm runs on real data, where no ground truth exists, exactly as it does on simulation.}}

\paragraph{Inner-loop budget and LLM backbone}
Each of the five evolved variants runs for the same total budget of 100 inner-loop iterations. The framework is LLM-agnostic in principle; in this case study we use a weighted pair of OpenAI models accessed through OpenRouter: \texttt{openai/gpt-4o} as the primary model (sampling weight 0.8) and \texttt{openai/gpt-4.1} as the secondary model (weight 0.2), with sampling parameters \texttt{temperature}\,=\,0.7, \texttt{top\_p}\,=\,0.95, and \texttt{max\_tokens}\,=\,8192. The same backbone is shared by all four GeoEvolve agents. Full hyperparameter settings, including the MAP-Elites population and GeoKnowRAG configuration, are reported in \ref{appendix:geoevolve_config}.

\subsection{\nv{Real-world Experiment}}\label{subsec:real_exp}

\subsubsection{\nv{Design}}\label{subsec:real_design}

\nv{Following Principle 2 of Module~\modThree (Section~\ref{subsec:module3}), the algorithm discovered on simulation is frozen and applied unchanged to each real-world dataset, with no further evolution, no hyperparameter retuning, and no knowledge retrieval on the real-world side. All algorithms in the real-world comparison are run with the same target number of regions $p$ (inferred from administrative units) and the same six-nearest-neighbor spatial weights matrix. \rv{Because each variant was evolved three times (Section~\ref{subsec:sim_results}), one program per variant must be carried forward to the real-world study. We select, for each variant, the run with the best simulation performance, and freeze that program; the selection uses simulation results only and no real-world data, so the real-world evaluation remains out-of-sample with respect to both evolution and selection. This is the best-of-$N$ protocol a practitioner would follow --- run the search a few times, keep the best algorithm --- and we report it explicitly because it means the real-world numbers reflect the best attainable program per variant rather than the average one. The corresponding distributions across all three runs are the ones given in Table~\ref{tab:baseline_simulation}.}}

\rv{\paragraph{Evaluation metrics} Following Principle 3 of Module~\modThree, the metrics used to report real-world performance are disjoint from the RandI fitness signal that drives the inner loop. Because regime discovery is a descriptive partitioning task (Section~\ref{subsec:problem_def}), the quantities below are model-fit statistics --- how much structural heterogeneity a partition accounts for --- computed by the evaluation harness identically for every algorithm, with the same target number of regions $p$, the same spatial weights, and the same data. Whether an algorithm recovers the \emph{correct} regime structure is established on the simulated data, where ground truth exists and agreement is measured by ARI (Section~\ref{subsec:sim_results}); the real-world study then asks whether the algorithm discovered there --- the best simulation run per variant, frozen with no further evolution or tuning --- transfers unchanged to data it has never seen.} Let $\{(X_i, Y_i)\}$ denote the dataset and $\hat{Y}_i=\mathcal{F}(X_i)$ the predictions of a candidate algorithm. We use four evaluation metrics. The \textbf{Sum of Squared Residuals (SSR)} measures absolute prediction error,
\begin{equation}
 \text{SSR}=\sum_{r=1}^R\sum_{i\in\mathcal{S}_r}(Y_i - \mathcal{F}_r(X_i))^2,
\end{equation}
where $\mathcal{F}_r$ is the regime-specific fit on region $\mathcal{S}_r$. The \textbf{SSR Reduction} ratio summarizes the proportional improvement of a multi-regime spatial model over a single global OLS fit:
\begin{equation}
 \text{SSR Reduction}=1-\frac{\sum_{r=1}^R\text{SSR}_{local,r}}{\text{SSR}_{global}}.
\end{equation}
The \textbf{Variance Ratio} reports the proportion of total variance in the target variable associated with between-region differences:
\begin{equation}
 \text{Variance Ratio}=\frac{\sigma^2_{between}}{\sigma^2_{within}+\sigma^2_{between}}.
\end{equation}
The \textbf{Root Mean Squared Error (RMSE)} reports cross-dataset prediction accuracy as the root mean squared deviation between predicted and observed $Y$. These four metrics are computed post-hoc and are never fed back into the evolution loop, so the LLM cannot game them.

\subsubsection{\nv{Datasets}}\label{subsec:real_datasets}

To evaluate the original and evolved versions of the spatial regime algorithm in a more realistic setting, we collected a set of empirical datasets covering both natural and socioeconomic topics. Table~\ref{tab:empirical_dataset} provides a summary of each dataset, including topic, source, target variable, and number of samples. The datasets vary in size (from hundreds to tens of thousands of samples) and span several domains, including climate, hydrology, politics, and health. Detailed descriptions, including the explanatory variables, are provided in \ref{appendix:details_datasets}.

\begin{table}[H]
 \scriptsize
 \centering
 \begin{tabular}{m{1.5cm}p{6cm}>{\centering\arraybackslash}m{4cm}>{\centering\arraybackslash}m{1.5cm}}
 \toprule
 \textbf{Topic} & \textbf{Dataset} & \textbf{Target Variable} & \textbf{\#Samples} \\
 \midrule
 Climate & ERA5-Land hourly data \citep{munoz2021era5} & Temperature of air at 2m above the surface & 83731 \\
 \hline
 Hydrology & Catchment Attributes and Meteorology for Large-sample Studies \citep{addor2017camels} & Mean daily discharge & 671 \\
 \hline
 Politics & \rw{US Presidential Election 2020} \citep{fotheringham2023measuring} & The percentage of votes for Democratic Party & 3108 \\
 \hline
 Health & PLACES: Local Data for Better Health \citep{greenlund2022places} & Prevalence of a type of disease & 2350 \\
 \bottomrule
 \end{tabular}
 \caption{An overview of real-world datasets, covering climate, hydrology, politics, and health.}
 \label{tab:empirical_dataset}
\end{table}

\subsection{\nv{Comparison Methods}}\label{subsec:comparison_methods}

\nv{Both the simulation experiment (Section~\ref{subsec:sim_exp}) and the real-world experiment (Section~\ref{subsec:real_exp}) compare the same ten algorithms under a unified protocol. The ten algorithms fall into two groups: six \emph{Two-stage K-Models variants}, which share an algorithmic ancestry with the seed and isolate the effect of LLM-driven evolution along a structured knowledge-injection design; and four \emph{classical regionalization baselines}, drawn from the broader spatial regime / regionalization literature, which examine whether any observed gains generalize beyond the K-Models family.}

\subsubsection{\nv{Six Two-stage K-Models variants}}\label{subsec:variants}

\nv{The six variants in this group share the Two-stage K-Models seed as their common starting point. The unevolved seed acts as the no-evolution lower bound. The remaining five variants span a $2{\times}3$ design that crosses the \emph{level} at which geographic knowledge enters the search (prompt-level vs.\ retrieval-level) with the \emph{specificity} of that knowledge (none / general / task-specific). Table~\ref{tab:variant_design} summarizes the design; the corresponding runtime configuration is given in~\ref{appendix:geoevolve_config}.}

\begin{table}[H]
 \scriptsize
 \centering
 \begin{tabular}{@{}l c c c c@{}}
 \toprule
 \textbf{Variant} & \textbf{Multi-agent} & \textbf{Prompt-level} & \textbf{Retrieval-level} & \textbf{Engine} \\
 & \textbf{dialogue} & \textbf{knowledge} & \textbf{knowledge} & \\
 \midrule
 2KModels (seed) & -- & -- & -- & seed only \\
 OpenEvolve (No GeoKnow.) & no & none & none & OpenEvolve \\
 OpenEvolve (General GeoKnow.) & no & general hint & none & OpenEvolve \\
 OpenEvolve (Specific GeoKnow.) & no & task-specific hint & none & OpenEvolve \\
 GeoEvolve (No GeoKnowRAG) & yes & analyzer-generated & disabled & GeoEvolve \\
 GeoEvolve & yes & analyzer-generated & GeoKnowRAG & GeoEvolve \\
 \bottomrule
 \end{tabular}
 \caption{Six Two-stage K-Models variants placed along the two-axis knowledge-injection design. ``Multi-agent dialogue'' indicates whether the Code Analyzer / Knowledge Retriever / Prompt Generator loop is active. ``Analyzer-generated'' indicates that the prompt content is produced at runtime by the Code Analyzer rather than written by hand. All five evolved variants run for the same total budget of $100$ inner-loop iterations.}
 \label{tab:variant_design}
\end{table}

\nv{This design allows us to decompose what would otherwise be a single ``GeoEvolve vs.\ seed'' comparison into two more informative sub-questions. The first asks whether multi-agent control is needed, or whether a static knowledgeable prompt already suffices. Comparing OpenEvolve (Specific GeoKnow.), which encodes task-specific geographic hints in a fixed prompt, against GeoEvolve (No GeoKnowRAG), which has the same agent dialogue but no curated knowledge corpus, isolates the contribution of the outer analyzer / prompt-generator loop. The second asks whether retrieval-level knowledge dominates prompt-level knowledge. Comparing all three OpenEvolve variants (no, general, or task-specific prompt) against the full GeoEvolve pipeline isolates the contribution of structured retrieval through GeoKnowRAG over static prompt-encoded hints. We hold the total evolutionary budget constant at $100$ inner-loop iterations across all five evolved variants, so any performance difference reflects the structure of the inner loop rather than search-budget disparities.}

\subsubsection{\nv{Four classical regionalization baselines}}\label{subsec:real_baselines}

\nv{A natural concern is that all six K-Models variants share the same algorithmic ancestry: the evolved variants are obtained by applying LLM-driven evolutionary search to the Two-stage K-Models seed. To examine whether the observed gains generalize beyond this seed family, we additionally compare against four representative algorithms drawn from the broader spatial regime / regionalization literature. The four baselines span distinct algorithmic paradigms developed over the past decades.}

\textbf{AZP} is \rw{the} regression-aware variant \rw{\citep{guo2023extending}} of the classical Automatic Zoning Procedure \rw{\citep{openshaw1977geographical,openshaw1995algorithms}} that performs local moves of border units between regions while preserving spatial contiguity and minimum-size constraints, accepting moves that reduce the total within-region SSR. \textbf{Region-K-Models} \citep{guo2023extending} is a partitioning method that iteratively reassigns each unit to the region whose local OLS fit best explains its observation, subject to spatial-neighbor and connectivity constraints. \textbf{GWR-Skater} \citep{helbich2013data} is a two-stage approach that first estimates location-specific regression coefficients via Geographically Weighted Regression and then clusters the resulting coefficient surface using SKATER, a minimum-spanning-tree-based regionalization algorithm. \textbf{Skater-reg} \citep{anselin2024endogenous} is an extension of SKATER that uses regression-fit residuals as the dissimilarity measure when cutting edges of the minimum spanning tree, directly aligning the partitioning objective with regression-fit quality.

\nv{The four classical baselines are matched to the K-Models variants on protocol (same target number of regions $p$, same six-nearest-neighbor spatial weights, same evaluation metrics) but differ in algorithmic family, so any performance gap reflects modeling-philosophy differences rather than implementation disparities.}

\section{\nv{Empirical Results: Simulated and Real-world Studies}}\label{sec:results}

\nv{This section reports the empirical results of running the ten algorithms (six Two-stage K-Models variants from Section~\ref{subsec:variants}, and four classical regionalization baselines from Section~\ref{subsec:real_baselines}) under the unified evaluation protocol of Section~\ref{sec:exp_design}. We begin in Section~\ref{subsec:what_modified} by introducing the algorithm discovered by SEGA as a new spatial regime model in its own right, comparing it with the Two-stage K-Models seed and tracing how the inner loop arrived at it over ten evolutionary rounds. Section~\ref{subsec:sim_results} then reports simulation performance on the full 150-grid evaluation set, and Section~\ref{subsec:real_results} reports the real-world results, focusing on GeoEvolve versus the four classical baselines; the detailed cross-variant comparison within the K-Models family on real-world data is provided in~\ref{appendix:realworld_variants}.}

\nv{\subsection{The Evolved Algorithm: GeoEvolve-2KModels}}\label{subsec:what_modified}

{
We refer to the algorithm produced by SEGA on the Two-stage K-Models seed as GeoEvolve-2KModels. This subsection introduces it as a new spatial regime model in its own right: how it differs structurally from the original 2KModels, how the inner loop arrived at those differences over ten evolutionary rounds, and what geographic-knowledge themes are embodied in the result.

\rv{\paragraph{Structural differences from the 2KModels seed} Compared with the original 2KModels, the discovered algorithm introduces two structural refinements, both visible in the round-by-round trace of Figure~\ref{fig:trajectory_comparison}. First, the \textbf{number of first-stage clusters} is no longer a fixed hyperparameter: where the seed always over-partitions into 20 clusters, the discovered algorithm evaluates a small candidate set ($2.5p$, $4p$, $\sqrt{n}$, and $p$ to $p+2$), runs each candidate through the merge stage, and keeps the partition with the lowest post-merge SSR --- turning a hand-set constant into an internal model-selection step. Second, it appends a \textbf{connectivity-preserving batch border-refinement stage}: after merging, border units are reassigned between adjacent regimes whenever the move lowers a spatially penalized SSR, subject to two connectivity checks that guarantee both the donor and the acceptor region remain contiguous, with early stopping under a patience rule bounding the added cost. Both refinements select internally by label-free SSR; neither involves any agreement-with-ground-truth quantity (Section~\ref{subsec:sim_design}).}

\begin{figure}[ht]
 \centering
 \includegraphics[width=\linewidth]{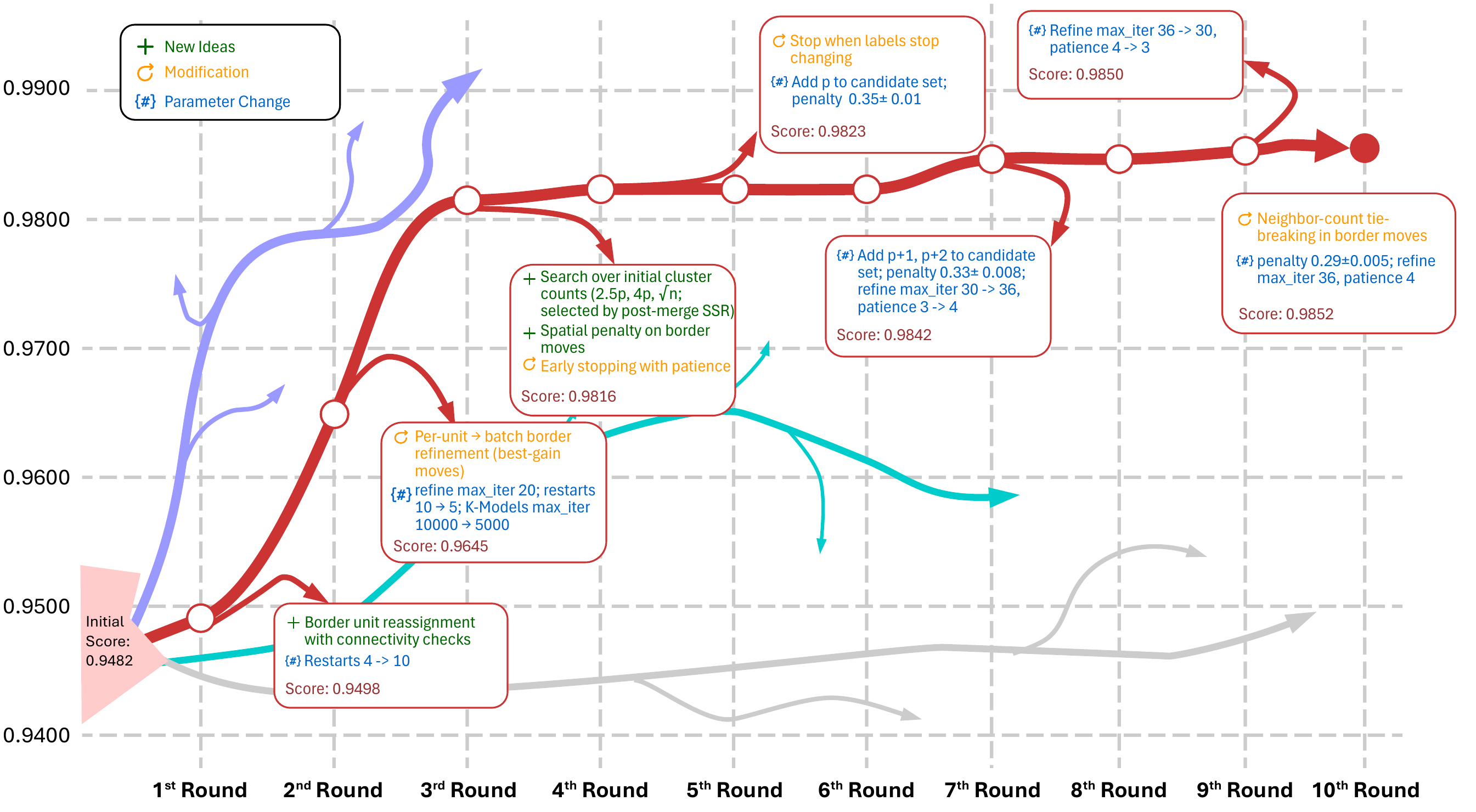}
 \caption{\rv{Round-by-round trajectory of the SEGA inner-loop search starting from the Two-stage K-Models seed, for the GeoEvolve (GeoKnowRAG) run reported in this paper. Annotated rounds add new ideas ($+$, green), code modifications ($\circlearrowleft$, orange), or parameter changes ($\{\#\}$, blue), with per-round scores marked in the figure. The plotted score is the inner-loop fitness: the mean Rand Index across the three grids used during evolution (Rectangular \#5, Voronoi \#75, Arbitrary \#149; Section~\ref{subsec:sim_design}), rising from $0.9482$ at the seed to $0.9852$ at round 10. The vertical axis is truncated because RandI is inflated by chance agreement on this task; the round-by-round changes, and the chance-corrected held-out results, are described in the text.}}
 \label{fig:trajectory_comparison}
\end{figure}

\paragraph{How the evolution unfolded over ten rounds} \rv{Figure~\ref{fig:trajectory_comparison} traces the inner-loop trajectory of the GeoEvolve (GeoKnowRAG) run across ten rounds, during which the mean Rand Index on the three training grids rises from $0.9482$ at the seed to $0.9852$. The gain is concentrated in the first three rounds, which contribute $+0.0334$ of the total $+0.0370$; the remaining seven rounds add $+0.0036$ between them, and three of those rounds (5, 6 and 8) accept no new variant at all. \textbf{Round 1} adds a border-unit reassignment step guarded by two connectivity checks, so that a unit only moves to a neighbouring region when both the donor and the acceptor remain spatially connected, and raises the number of restarts from 4 to 10; this operationalizes the geographic principle that adjacent units are more likely to share a regime than distant ones (Tobler's first law \citep{tobler1970computer}) as a local, topology-preserving move rather than a global re-partition. \textbf{Round 2} converts that step from per-unit to batch reassignment, evaluating all candidate moves and applying the best-gain set in one pass, and lifts the score to $0.9645$ --- the single largest jump before round 3. \textbf{Round 3} makes the one change that alters the algorithm's search space: the number of first-stage clusters, fixed at 20 in the seed, becomes a small search over $2.5p$, $4p$ and $\sqrt{n}$, each run through the merge stage and selected by SSR; the same round adds a spatial penalty on border moves and early stopping with patience, and the score reaches $0.9816$. Rounds 4--10 tune this machinery rather than extend it: the candidate set is widened to include $p$, $p+1$ and $p+2$, the spatial penalty is annealed from $0.35$ to $0.29$ with a small deterministic jitter for diversity across restarts, and the refinement budget is adjusted between 30 and 36 iterations with patience 3--4, ending at $0.9852$. The end-of-trajectory algorithm is the GeoEvolve-2KModels reported in the remainder of this paper. Two cautions apply to this figure. First, the trajectory score is the mean RandI over the three \emph{training} grids \#5, \#75 and \#149 and is therefore not comparable with the per-pattern values in Table~\ref{tab:baseline_simulation}, which are computed on the 150-grid pool; the same algorithm scores $0.9634$ RandI there. Second, RandI is the optimization signal rather than an independent measure of quality, and it is inflated by chance agreement on this task, which is why Table~\ref{tab:baseline_simulation} reports the chance-corrected ARI as the agreement metric.} \nv{A complete round-by-round trace of this evolutionary process, including the full source code at each round, the LLM-generated edits, and the per-round fitness, is provided as an interactive web page, linked from the project page (\url{https://peng-luo.com/sega/}), as a supplement to Figure~\ref{fig:trajectory_comparison}.}

\rv{\paragraph{Geographic-knowledge themes embodied in the evolved algorithm} Across these refinements, three geographic-knowledge themes are repeatedly invoked. On \textbf{spatial contiguity}, every border-unit move is guarded by two connectivity checks that guarantee both the donor and the acceptor region remain spatially connected --- the algorithm treats contiguity as a hard constraint to be preserved through every local improvement, not repaired afterwards. On \textbf{local spatial dependence}, the border-refinement stage operationalizes Tobler's first law \citep{tobler1970computer}: reassignments are confined to units on regime boundaries, where a unit's regime membership is most uncertain and most informed by its neighbours, and a spatial penalty on border moves discourages jagged, noise-driven boundaries. On \textbf{scale as a modelling choice}, the search over the number of first-stage clusters ($2.5p$, $4p$, $\sqrt{n}$) treats the initial partition scale not as a fixed hyperparameter but as a quantity to be selected against the data by post-merge SSR, echoing the sensitivity of regionalization outcomes to the chosen unit scale. The broader interpretation of these themes for autonomous geospatial algorithm discovery is taken up in Section~\ref{subsec:insights_complementarity}.}
}

\subsection{\nv{Simulation Results}}\label{subsec:sim_results}

\rv{\paragraph{Performance on the simulated grids} The K-Models family outperforms the four classical regionalization baselines by a wide margin under the same protocol (Table~\ref{tab:baseline_simulation}): the baselines' average SSR ranges from ${\approx}206$ (AZP) to ${\approx}403$ (Skater-reg) against ${\approx}68$--$87$ for the family, and their ARI from $0.330$ to $0.624$ against $0.806$--$0.869$ --- a separation an order of magnitude larger than any run-to-run spread, and consistent with the fact that the simulated data follow a piecewise-linear regime model whose assumption 2KModels and its descendants share but methods clustering GWR coefficient surfaces or attribute distances do not directly optimize. Within the evolved family, averaged over three independent runs, GeoEvolve with GeoKnowRAG attains the best three-pattern averages --- SSR $67.53 \pm 18.24$, ARI $0.869 \pm 0.027$, a $14.9\%$ SSR reduction and a $+0.057$ ARI gain over the seed --- but it does not dominate every pattern (GeoEvolve without GeoKnowRAG has the lower SSR on Voronoi and Arbitrary), and the effect of geographic knowledge is non-monotone: OpenEvolve without geographic knowledge is worse than the seed it started from (SSR $86.64$ versus $79.38$; ARI $0.806$ versus $0.812$), a general knowledge prompt improves SSR to $72.95$, a task-specific one only to $78.30$. Because the run-to-run standard deviations ($5.32$--$18.24$ on average SSR) are of the same order as the gaps between variants ($67.53$--$86.64$), we report the ordering among evolved variants as descriptive rather than as evidence that one is reliably better than another. Complementing this run-to-run view, the discovered algorithm itself is reliably better than the seed on the evaluation pool: a paired Wilcoxon signed-rank test over the 150 grids gives a median SSR improvement of 15.0 ($p < 10^{-23}$) and a median ARI gain of 0.080 ($p < 10^{-17}$). Finally, the runtime column shows that the cost of the accuracy gain is not intrinsic: the evolved algorithms range from $0.91$ to \rw{$10.45$} seconds per grid\rw{, excluding one run without early stopping that averages $188.9$~s}, an ordering that follows how much search each performs at inference time rather than how accurate it is; full token, \rw{elapsed-time} and inference-time accounting is given in \ref{appendix:cost}. \rw{A matched-budget control confirms that these gains are not an artifact of additional search effort. Giving the unmodified seed the same cluster-count search with random restarts, under the evolved algorithm's own per-grid time budget, does narrow the SSR gap --- SSR being the quantity any such search optimizes directly --- but still leaves it short on \emph{recovery of the true regime structure}, which is what the task is defined on and which the comparison should therefore be read on (ARI $0.854$ versus $0.894$, $p = 2.3\times10^{-10}$); one-component ablations locate the advantage in the border-refinement stage and its spatial penalty rather than in the cluster-count search --- the penalty being a component whose value is \emph{negative} in the algorithm's own selection objective, which no tuning of the seed's parameters against that objective could have reached (\ref{appendix:equalbudget}).}}

\begin{figure}[H]
 \centering
 \includegraphics[width=\linewidth]{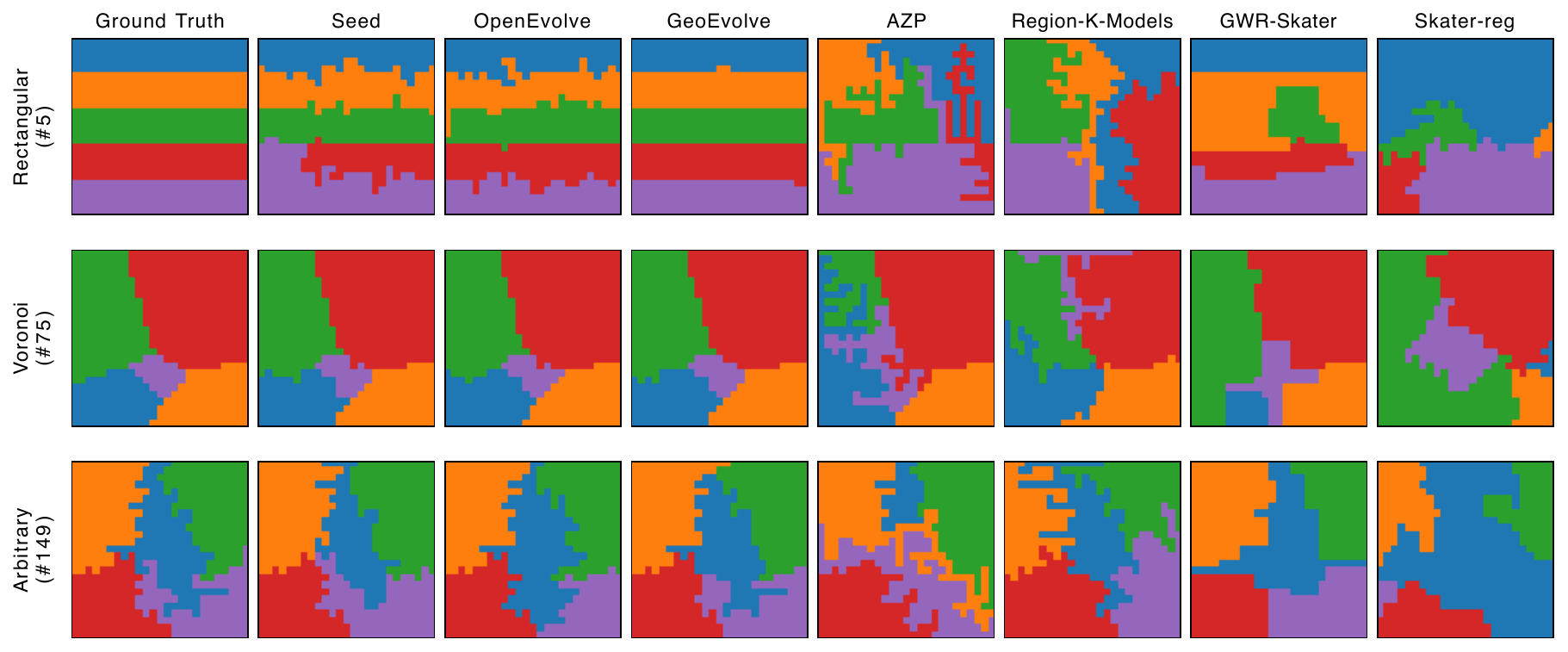}
 \caption{Region delineation results on the three grids used during SEGA evolution, Rectangular \#5 (top row), Voronoi \#75 (middle row), Arbitrary \#149 (bottom row). The first column shows the ground-truth spatial regime. The next three columns show reconstructions from the seed (Two-stage K-Models, ``Seed'') and \rv{two representative evolved variants} (OpenEvolve with general geographic-knowledge prompt and GeoEvolve with full GeoKnowRAG; see Table~\ref{tab:baseline_simulation}). The last four columns show reconstructions from the four classical regionalization baselines (AZP, Region-K-Models, GWR-Skater, Skater-reg). All algorithms are evaluated on the same grid under the unified protocol of Section~\ref{sec:exp_design}; colors are aligned across columns using Hungarian assignment so that the same region in different columns shares a color where possible.}
 \label{fig:simulation_examples}
\end{figure}

Figure~\ref{fig:simulation_examples} illustrates the reconstructed spatial regimes on the three grids used during SEGA evolution (Rectangular \#5, Voronoi \#75, Arbitrary \#149), comparing the seed, the two best evolved variants, and the four classical regionalization baselines on the same grids. \nv{Within the K-Models family,} both evolved versions of 2KModels produce regions that resemble the ground truth, while the original 2KModels exhibits some misidentified boundaries and less coherent regions; the GeoEvolve-evolved version is the closest to the ground truth across all three patterns. \nv{The four classical baselines reveal a sharper qualitative gap: AZP and Region-K-Models produce visibly fragmented partitions, particularly on the Rectangular grid where the ground-truth structure is the most regular; GWR-Skater preserves the main horizontal structure on Rectangular but loses coherence on Voronoi and Arbitrary; Skater-reg merges the central regions and reduces the partition to two or three dominant regions on all three patterns. These visual differences are consistent with the SSR gap reported in Table~\ref{tab:baseline_simulation}, which the K-Models family closes by directly optimizing within-region regression fit.}

\rv{\begin{table}[ht]
    \tiny
    \centering
    \setlength{\tabcolsep}{3pt}
    \resizebox{\linewidth}{!}{%
    \begin{tabular}{@{}l ccc c ccc c r@{}}
    \toprule
    \multirow{2}{*}{\textbf{Algorithm}} & \multicolumn{4}{c}{\textbf{SSR} $\downarrow$} & \multicolumn{4}{c}{\textbf{ARI} $\uparrow$} & \multirow{2}{*}{\textbf{Time (s)}} \\
    \cmidrule(lr){2-5}\cmidrule(lr){6-9}
    & \textbf{Rect.} & \textbf{Voronoi} & \textbf{Arb.} & \textbf{Avg.} & \textbf{Rect.} & \textbf{Voronoi} & \textbf{Arb.} & \textbf{Avg.} & \\
    \midrule
    Skater-reg                               & 476.82 & 342.18 & 390.66 & 403.22 & 0.253 & 0.391 & 0.347 & 0.330 & 5.46 \\
    Region-K-Models                          & 388.13 & 295.68 & 355.04 & 346.28 & 0.348 & 0.457 & 0.429 & 0.411 & 0.02 \\
    GWR-Skater                               & 215.31 & 204.66 & 288.54 & 236.17 & 0.613 & 0.691 & 0.567 & 0.624 & 6.83 \\
    AZP                                      & 237.92 & 171.73 & 208.32 & 205.99 & 0.435 & 0.531 & 0.511 & 0.492 & 0.59 \\
    OpenEvolve (No GeoKnow.)                 & 80.92 $\pm$ 11.08 & 79.42 $\pm$ 10.61 & 99.57 $\pm$ 16.76 & 86.64 $\pm$ 12.79 & 0.804 $\pm$ 0.006 & 0.860 $\pm$ 0.015 & 0.754 $\pm$ 0.015 & 0.806 $\pm$ 0.010 & 2.99 $\pm$ 1.85 \\
    2KModels (seed)                          & 75.46 & 72.29 & 90.38 & 79.38 & 0.804 & 0.863 & 0.770 & 0.812 & 0.67 \\
    OpenEvolve (Specific GeoKnow.)           & 73.38 $\pm$ 4.63 & 71.60 $\pm$ 5.15 & 89.91 $\pm$ 6.58 & 78.30 $\pm$ 5.32 & 0.820 $\pm$ 0.016 & 0.876 $\pm$ 0.015 & 0.769 $\pm$ 0.019 & 0.821 $\pm$ 0.016 & 7.42 $\pm$ 8.53 \\
    OpenEvolve (General GeoKnow.)            & 66.69 $\pm$ 9.16 & 66.96 $\pm$ 9.34 & 85.20 $\pm$ 14.41 & 72.95 $\pm$ 10.66 & 0.843 $\pm$ 0.038 & 0.891 $\pm$ 0.028 & \textbf{0.799 $\pm$ 0.033} & 0.844 $\pm$ 0.030 & 0.91 $\pm$ 0.45 \\
    GeoEvolve (No GeoKnowRAG)                & 67.43 $\pm$ 10.56 & \textbf{63.17 $\pm$ 9.06} & \textbf{75.16 $\pm$ 15.49} & 68.58 $\pm$ 11.64 & 0.806 $\pm$ 0.015 & 0.883 $\pm$ 0.017 & 0.780 $\pm$ 0.021 & 0.823 $\pm$ 0.017 & 1.68 $\pm$ 1.60 \\
    \textbf{GeoEvolve (GeoKnowRAG)}          & \textbf{55.60 $\pm$ 1.60} & 63.64 $\pm$ 15.41 & 83.35 $\pm$ 39.77 & \textbf{67.53 $\pm$ 18.24} & \textbf{0.909 $\pm$ 0.083} & \textbf{0.898 $\pm$ 0.031} & \textbf{0.799 $\pm$ 0.049} & \textbf{0.869 $\pm$ 0.027} & \rw{69.95 $\pm$ 103.05 ($10.45^{\dagger}$)} \\
    \bottomrule
    \end{tabular}}
    \caption{Simulation comparison of four classical regionalization baselines, the unevolved Two-stage K-Models seed, and the five evolved variants on the 150 high-noise grids (50 per pattern). All metrics are computed externally by the evaluation harness (Section~\ref{subsec:sim_design}); each evolved row is the mean $\pm$ SD over three independent evolution runs, while the seed and the baselines are deterministic given the data. ARI is the chance-corrected agreement metric; RandI serves only as the inner-loop fitness signal (Figure~\ref{fig:trajectory_comparison}). Time is the mean runtime per 625-cell grid (\ref{appendix:cost})\rw{; for GeoEvolve (GeoKnowRAG), the value in parentheses is the trimmed mean$^{\dagger}$ excluding one run whose refinement step lacks early stopping}. Rows sorted from worst to best by average SSR; bold = best within column.}
    \label{tab:baseline_simulation}
\end{table}}

\subsection{\nv{Real-world Results}}\label{subsec:real_results}
\begin{figure}[H]
 \centering
 \includegraphics[width=\linewidth]{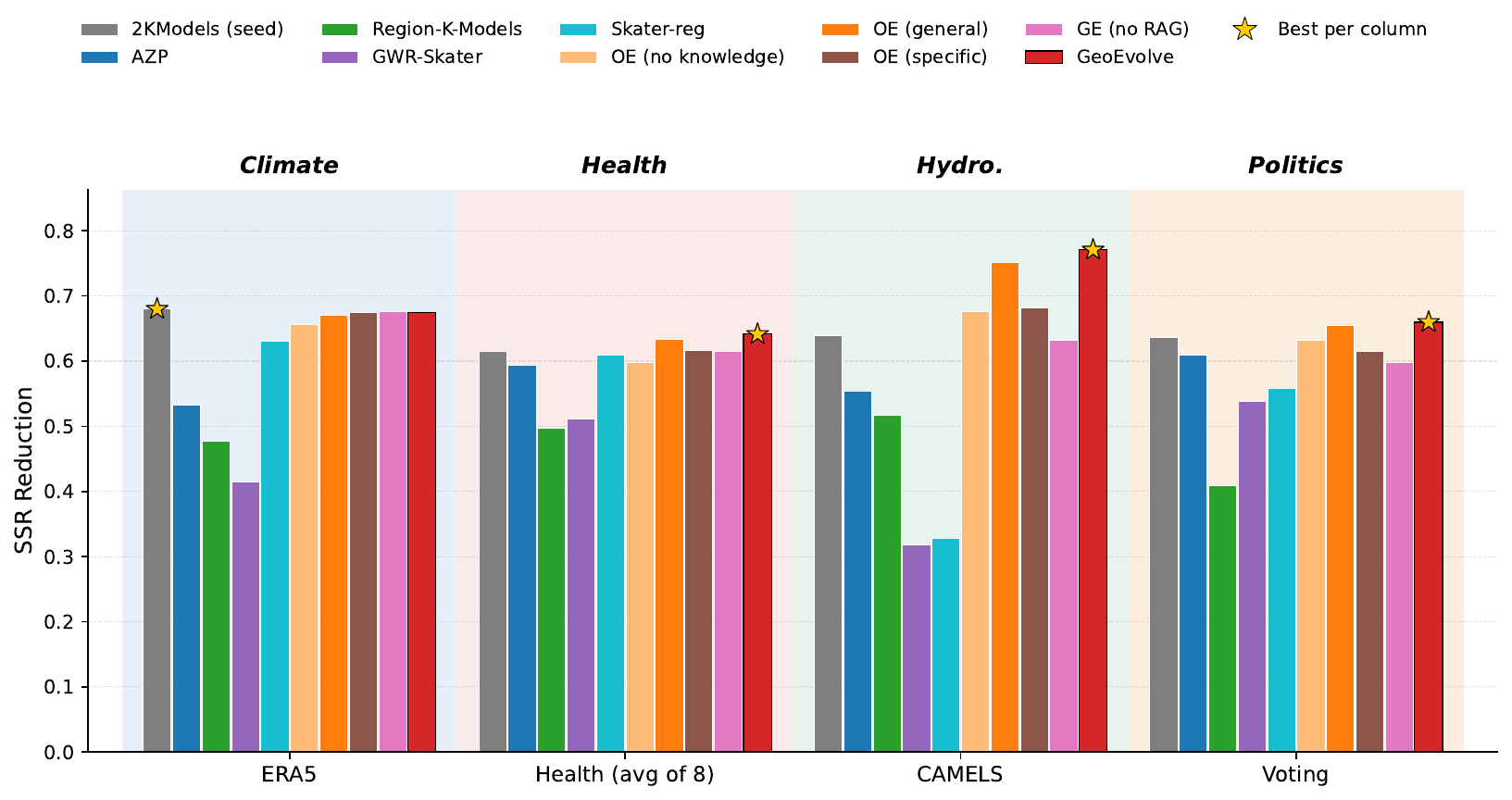}
 \caption{\nv{Ten-algorithm SSR-Reduction comparison across the four real-world domains, with one column per domain (Climate, Health, Hydro., Politics) on tinted backgrounds. The Health column reports the mean over the eight Health datasets (Arthritis, BPHigh, Cancer, Asthma, Depression, Diabetes, Obesity, Stroke); the other three domains each contain one dataset (ERA5, CAMELS, Voting). The ten algorithms are the unevolved Two-stage K-Models seed (grey), the four classical regionalization baselines (AZP, Region-K-Models, GWR-Skater, Skater-reg), the three OpenEvolve variants (no / general / task-specific GeoKnow.\ prompt), GeoEvolve without GeoKnowRAG, and the full GeoEvolve with GeoKnowRAG (red, highlighted with a darker outline). Gold stars mark the best algorithm per column. The corresponding cross-dataset means on the four real-world metrics are reported in Table~\ref{tab:baseline_realworld_summary}; the per-dataset breakdown on all 11 real-world datasets and the corresponding regime maps are provided in~\ref{appendix:realworld_variants}.}}
 \label{fig:baseline_comparison_real_world}
\end{figure}

\nv{We now turn from simulated data, where the ground-truth partition is known and the fitness signal can be evaluated directly, to real-world data, where neither is available and the discovered algorithm is asked to generalize out of sample. We apply the frozen GeoEvolve algorithm to the 11 real-world datasets and compare it against the unevolved Two-stage K-Models seed and the four classical regionalization baselines under the unified evaluation protocol of Section~\ref{subsec:real_design}. Figure~\ref{fig:baseline_comparison_real_world} reports a compact ten-algorithm SSR-Reduction summary by domain (which also includes the three OpenEvolve variants and GeoEvolve without GeoKnowRAG), and Table~\ref{tab:baseline_realworld_summary} reports cross-dataset means on the four quality metrics for the six-algorithm main comparison (seed, four classical baselines, GeoEvolve). The detailed within-family comparison of all six K-Models variants on real-world data (per-dataset, four-metric breakdown, with corresponding regime maps and dominant-covariate maps) is provided in \ref{appendix:realworld_variants}. Compared with the simulation, the real-world picture is more nuanced.}

GeoEvolve attains the lowest cross-dataset average RMSE (1.220) and the highest SSR Reduction (0.658), with nine first-place wins on SSR Reduction out of 11 datasets, six wins on average $R^2$, and five wins on RMSE, more than any other algorithm on all three metrics; on Variance Ratio it has the highest mean (0.482) and four wins (second to Skater-reg's five). AZP reaches the highest mean $R^2$ (0.830) and is the most competitive classical baseline on per-dataset $R^2$ and RMSE wins (4 each out of 11), illustrating that older regionalization methods remain competitive on individual datasets when the underlying regimes are nearly piecewise constant in the response variable. The unevolved 2KModels seed (mean $R^2$ 0.816, mean SSR Reduction 0.625) holds its own as a strong baseline, ahead of all four classical baselines on SSR Reduction. Among the four classical regionalization baselines, three (Region-K-Models, GWR-Skater, Skater-reg) fall noticeably behind on SSR Reduction (0.488--0.582 vs.\ 0.658 for GeoEvolve and 0.625 for the seed).

Two qualitative conclusions follow. First, \rv{an algorithm discovered entirely on simulated grids transfers to real data without modification}: GeoEvolve is competitive with, or better than, every classical regionalization baseline on real-world data, despite the simulator being matched to the K-Models modeling assumption. \rv{The claim here is about transfer of the algorithm, not about predictive generalization to unobserved locations: these are fit statistics on the datasets being partitioned (Section~\ref{subsec:real_design}), and what they show is that a procedure developed against a synthetic regime model remains well behaved on empirical data whose generating process it was never shown.} Second, no single algorithm dominates across all metrics and all datasets: AZP is unexpectedly strong on $R^2$, Skater-reg wins the most datasets on Variance Ratio despite a lower mean, and GeoEvolve provides the tightest regression fit, the largest SSR Reduction, and the highest mean Variance Ratio. This suggests that performance is metric-dependent and that practitioners should consider their downstream use case when selecting an algorithm.

\begin{figure}[H]
 \centering
 \includegraphics[width=\linewidth]{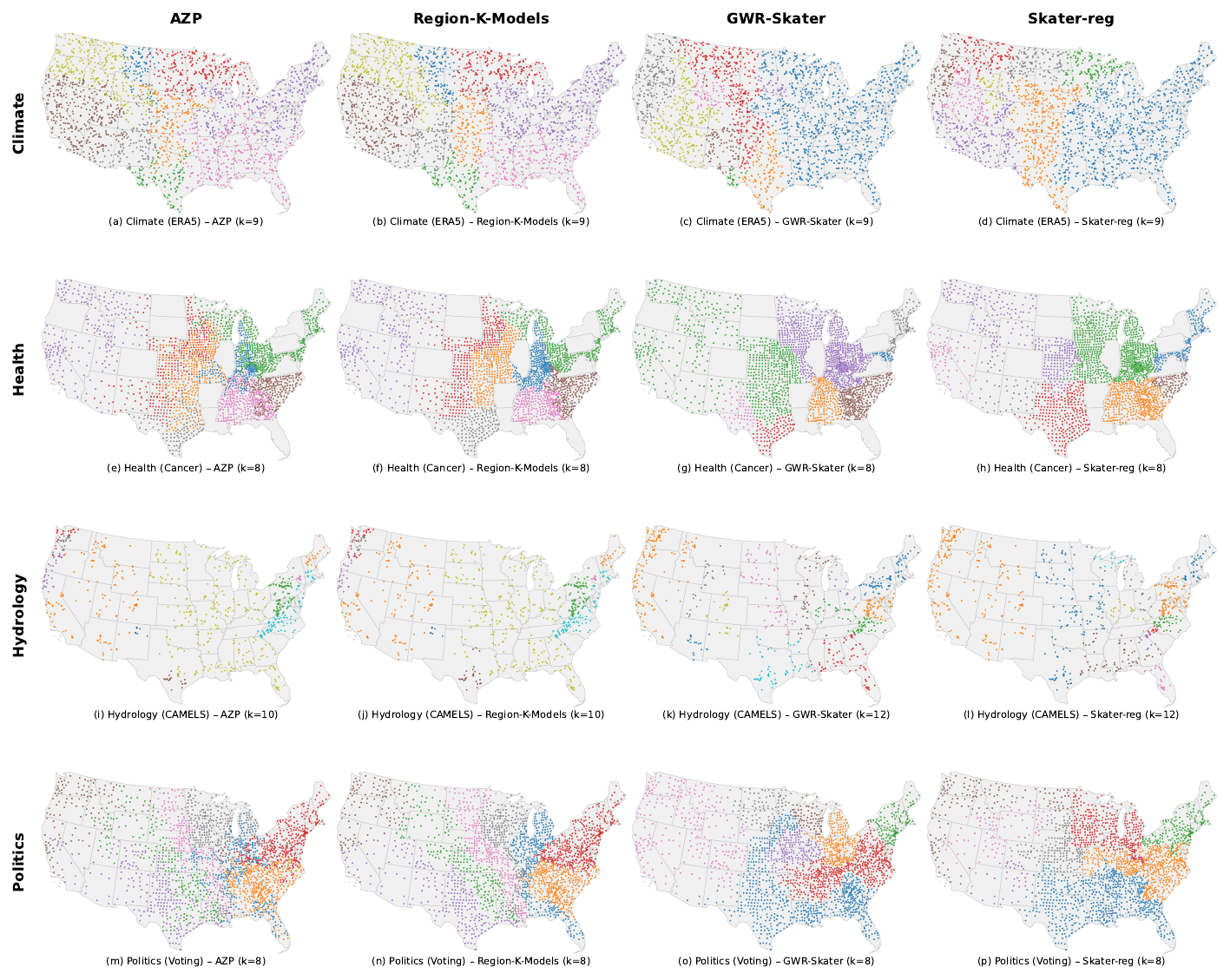}
 \caption{Regime-partitioning maps produced by the four classical baselines on four representative real-world datasets (Climate / Health / Hydrology / Politics). Classical regionalization baselines tend to produce more spatially fragmented partitions, particularly in the politics and health domains. The corresponding K-Models family regime maps are provided in \ref{appendix:realworld_variants} (Figure~\ref{fig:evolved_regime_maps}).}
 \label{fig:baseline_regime_maps}
\end{figure}

The spatial structure underlying these metric differences is visible in the regime-partitioning maps. Figure~\ref{fig:baseline_regime_maps} shows the partitions produced by the four classical baselines on four representative datasets (Climate / Health / Hydrology / Politics). The four classical baselines tend to produce more spatially fragmented partitions, particularly in the politics and health domains. The corresponding regime maps for the six K-Models family variants (and a fine-grained cross-variant breakdown) are provided in \ref{appendix:realworld_variants} (Figure~\ref{fig:evolved_regime_maps}); they generate more spatially coherent regimes with clearer boundaries, with GeoEvolve being the most coherent.

\begin{table}[H]
 \scriptsize
 \centering
 \begin{tabular}{@{}l rrrr@{}}
 \toprule
 \textbf{Algorithm} & \textbf{Avg.\ $R^2$} $\uparrow$ & \textbf{Avg.\ RMSE} $\downarrow$ & \textbf{SSR Red.} $\uparrow$ & \textbf{Var.\ Ratio} $\uparrow$ \\
 \midrule
 2KModels (seed) & 0.816 & 1.273 & 0.625 & 0.465 \\
 \midrule
 \multicolumn{5}{@{}l}{\textbf{Classical regionalization baselines}}\\
 AZP & \textbf{0.830} & 1.228 & 0.586 & 0.373 \\
 Region-K-Models & 0.789 & 1.440 & 0.489 & 0.344 \\
 GWR-Skater & 0.800 & 1.303 & 0.488 & 0.358 \\
 Skater-reg & 0.812 & 1.335 & 0.582 & 0.395 \\
 \midrule
 \textbf{GeoEvolve} & 0.825 & \textbf{1.220} & \textbf{0.658} & \textbf{0.482} \\
 \bottomrule
 \end{tabular}
 \caption{Real-world summary: cross-dataset means on the four quality metrics. Bold = best within column. The detailed within-family breakdown of all six K-Models variants (the seed, three OpenEvolve variants, and GeoEvolve with and without GeoKnowRAG), with per-dataset and per-metric values, is given in Figure~\ref{fig:baseline_comparison_real_world_full} (\ref{appendix:realworld_variants}).}
 \label{tab:baseline_realworld_summary}
\end{table}

\section{\nv{Discussion: Toward Autonomous Geospatial Algorithm Discovery}}\label{sec:discussion}
\nv{In this section we step back and interpret what the case study suggests about the SEGA framework introduced in Section~\ref{sec:methods}. Building on the structural description of the discovered algorithm given in Section~\ref{subsec:what_modified}, Section~\ref{subsec:insights_complementarity} asks what general lessons can be drawn from those refinements and how SEGA complements expert-driven design. Section~\ref{subsec:limitations} outlines limitations and risks.}

\subsection{\nv{Insights and Complementarity with Expert-driven Design}}\label{subsec:insights_complementarity}
{
Taken together, the case study supports a more general suggestion: \emph{when the optimization objective for a geospatial problem can be specified in a computable form, SEGA can be a useful complement to traditional, expert-driven algorithm development.} In the spatial regime case, the objective of reducing within-region structural variability while increasing between-region differentiation was encoded as the bounded, ground-truth-anchored Rand Index. Within this setting, the multi-agent system surfaced algorithmic refinements \rv{(an internal search over the initial partition scale, connectivity-preserving border refinement, spatially penalized reassignment)} that align with established geographic intuitions and are useful starting points for further analysis. Table~\ref{tab:potential_problems} outlines several other geospatial problems for which optimization objectives can plausibly be formalized and which may therefore benefit from a similar treatment.

\begin{table}[H]
 \scriptsize
 \centering
 \begin{tabular}{p{3cm}p{5cm}p{5cm}}
 \toprule
 \textbf{Geospatial Problem} & \textbf{Formalizable Objective} & \textbf{Possible Direction for LLM-Assisted Exploration} \\
 \midrule
 Spatial interpolation & Minimize prediction error at unobserved locations (e.g., cross-validated RMSE) & Search over interpolation strategies, kernel functions, neighborhood structures \\
 \hline
 Spatial clustering & Increase spatial coherence and attribute homogeneity (e.g., silhouette score + contiguity) & Evolve partitioning strategies, distance metrics, merge criteria \\
 \hline
 Spatial prediction with uncertainty & Reduce prediction interval width while maintaining coverage (e.g., interval score) & Explore variants of conformal prediction and calibration methods \\
 \hline
 Change detection & Improve detection accuracy under spatial and temporal constraints (e.g., F1-score) & Evolve feature extraction, thresholding, and post-processing pipelines \\
 \bottomrule
 \end{tabular}
 \caption{Examples of geospatial problems with formalizable objectives that may be amenable to SEGA-style algorithm discovery. In each case, the most demanding intellectual work remains the formulation step.}
 \label{tab:potential_problems}
\end{table}

SEGA is best understood as \emph{complementary} to expert-driven design rather than a replacement. Conceptual innovations from researchers, adding spatial weights to a regression model, embedding contiguity constraints, introducing new estimators, continue to mark out the design space and define what counts as a good answer; LLM-assisted search then fills in parts of that space that are tedious to explore manually. The practical value of SEGA comes less from any single ``intelligent'' move and more from two properties LLMs supply at low marginal cost: \emph{diversity} (many candidate structural modifications generated and tried within a short time window) and \emph{iteration speed} (each round of variation, evaluation, and selection completing in minutes rather than weeks). \rv{The combination of an internal model-selection step over the initial partition scale with a connectivity-guarded batch border refinement assembled in our evolved 2KModels is one such combination} that no individual researcher would have necessarily prioritized in advance, but that the search procedure surfaced by trying many alternatives. The implication for GIScience is that the bottleneck in geospatial algorithm development is shifting from \emph{implementation effort} toward \emph{specification effort}: the geographer's role takes on additional dimensions as a problem architect who defines the objective landscape in which the LLM-driven search operates, while classical spatial algorithms remain valuable both as seeds for evolution and as baselines that anchor the comparison.}

\rv{\subsection{Shared Infrastructure: Benchmarks, Traces, and Validation}\label{subsec:infrastructure}
Making algorithm discovery a normal mode of geospatial research, rather than a series of one-off demonstrations, requires shared infrastructure for evaluating what the searches produce. The central object is a common benchmark: a fixed interface (input/output signature), a fixed evaluation protocol in which metrics are computed by the harness rather than by the candidate, and an explicit split between the instances a search may optimize against and those reserved for reporting. Validation then happens on that benchmark: evolved algorithms should be compared against human-designed algorithms of the same class --- not only against their own seed --- reported as distributions over independent evolution runs rather than single trajectories, and finally evaluated on data the search never touched. Seed algorithms and injected domain knowledge, by contrast, should \emph{not} be standardized: they are the researcher's contribution, the analogue of a hypothesis, and different seeds and knowledge bases exploring the same benchmark is precisely the diversity that makes shared evaluation informative. The remaining piece is a standard evolution-trace format, recording per round the candidate program, its fitness, and the prompt or knowledge that produced it. Beyond reproducibility, traces make credit assignable --- a discovery is jointly produced by whoever contributed the seed, the knowledge corpus, and the objective and search design --- and they are research material in their own right: with a common format, questions such as which prompting regimes produce structural innovation rather than parameter tuning become empirical questions answerable across laboratories.

\subsection{Beyond Single-Objective, Simulation-Friendly Tasks}\label{subsec:beyond_single_objective}
Extending SEGA past single-objective tasks requires distinguishing two situations. When several objectives must all be respected but no trade-off curve is needed (e.g., accuracy alongside runtime), lightweight randomized scalarization suffices: recent LLM-driven evolution systems define several fitness metrics and randomly select one per generation \citep{li2026multiagent}. When the trade-off curve itself is the object of interest --- within-region homogeneity against region-size balance, or accuracy against uncertainty-interval width --- the search must instead maintain a set of non-dominated alternatives, via explicit non-dominated sorting \citep{deb2000nsga} or a quality-diversity archive indexed by the competing objectives \citep{mouret2015illuminating}; the MAP-Elites feature map already used here for \emph{(score, complexity)} is a natural place for that substitution. Characterizing which geospatial problems need a frontier rather than a compromise is, to us, the most concrete piece of future work.}

\rw{\subsection{Toward Trustworthy and Responsible Self-Evolving GeoAI}\label{subsec:requirements}
Self-evolution sharpens a known ethical problem of spatial analysis. \citet{chen2026ethical} show that endogenous bias arises whenever a global optimization goal diverges from the local needs of particular regions or communities: the aggregate looks sound while some places or groups are systematically disadvantaged. An evolutionary search is an aggregate optimizer by construction, keeping any variant that raises mean fitness, even one that concentrates errors in a few regions, and since no one designed the evolved mechanism, such trade-offs are easily missed. Three checkpoints follow. \textbf{Before the search}, the problem should be deliberately delimited, and where distributional consequences matter, the objective should register them (e.g., worst-region terms alongside the mean, Section~\ref{subsec:beyond_single_objective}) rather than average them away; our restriction of regime discovery to descriptive partitioning (Section~\ref{subsec:problem_def}) is a modest instance. \textbf{Before deployment}, the evolved algorithm should be diagnosed, not only scored: visual analytics \citep{chen2026ethical} can map its errors and uncertainty, which are rarely uniform over space and so translate into spatially uneven risk, and check the spatial continuity of the parameters it estimates. \textbf{At deployment}, performance established in one study region cannot be assumed to transfer, so local diagnostics, recalibration where needed, and disaggregation of results by spatial unit and population group should precede any decision. However much of the search is delegated, the judgement of whom its results serve remains the geographer's.}

\subsection{\nv{Limitations and Risks}}\label{subsec:limitations}
{While the present results are encouraging, several limitations and risks deserve attention. \textbf{Dependence on objective-function quality.} The hypothesis that problems with well-defined objectives can be partly addressed by autonomous search places a heavy burden on the objective itself. Poorly specified loss functions may lead the evolutionary system to favor algorithms that optimize the metric while drifting away from the intended geographic phenomenon; the choice of RandI as the fitness signal in our case study is a concrete instance of how carefully this choice must be made. \textbf{Problems with ill-defined objectives.} Not all geospatial problems admit clean, easily formalizable objectives. Tasks such as cartographic generalization, place characterization, or qualitative spatial reasoning involve aesthetic, cognitive, or contextual judgements that are difficult to reduce to a single scalar metric; for such problems, hybrid approaches combining human judgement with autonomous search are likely to be needed. \textbf{Interpretability and trust.} Evolved algorithms can be more difficult to interpret than manually designed ones; \rv{the evolved 2KModels' interacting heuristics --- a candidate-set search over initial cluster counts nested inside the two-stage procedure, and a border-refinement stage whose penalty magnitude, iteration budget and patience were tuned by the search rather than derived from theory --- lack the conceptual transparency of methods derived from first principles}, which raises legitimate concerns for scientific reproducibility. \textbf{Knowledge-base bias.} The geographic knowledge base $\mathcal{K}$ inevitably reflects the choices, blind spots, and disciplinary tendencies of its source materials; over-representing certain methodological traditions (e.g., spatial econometrics) while under-representing others (e.g., qualitative geography) can steer the evolutionary search toward a narrow region of the solution space. \rw{\textbf{Single seed and task family.} All empirical evidence comes from one seed algorithm in one task family: the case study demonstrates feasibility, not generality. Establishing SEGA as a general paradigm requires demonstrations across algorithm families under the same evaluation discipline; until then, framework-level statements should be read as scoped to problems resembling the one studied here.} \textbf{Computational cost and reproducibility.} LLM-driven evolution requires substantial computational resources and depends on language models whose behavior may change across versions, making bit-for-bit reproducibility difficult.}

\section{\nv{Conclusion}}\label{sec:conclusion}
{This paper presented \emph{Self-Evolving Geospatial Algorithms} (SEGA), a framework for developing geospatial algorithms through a structured, evaluation-driven LLM-assisted inner loop alongside the traditional expert-driven discovery carried out by researchers. SEGA rests on three required modules: a two-loop architecture with embedded geographic knowledge (adopted from GeoEvolve), an automatic code formatter that turns any raw seed algorithm into the standardized input the inner loop requires, and a simulation-driven training protocol that disciplines the use of data and metrics so that real-world performance counts as out-of-sample evidence. The spatial regime case study validates the full stack: the knowledge-guided evolved Two-stage K-Models outperforms both its seed and knowledge-free baselines on 150 simulated grids, transfers without retuning to 11 real-world datasets across climate, health, hydrology, and politics, and remains competitive with four classical regionalization baselines on real data. \rw{We close with a hypothesis that this case study supports but cannot yet establish: when a geospatial problem admits a computable optimization objective, much of the inner-loop work of algorithm refinement may be delegable to LLM-driven systems; demonstrating this beyond a single seed and task family is the necessary next step. What the study does show is the geographer's contribution shifting upstream, toward articulating new questions, designing meaningful loss functions, and building trustworthy evaluation protocols.}}

\section{Data and Code Availability}

\nv{The complete code and data accompanying this paper, together with an interactive trace of the evolution it reports, are available from the project page: \url{https://peng-luo.com/sega/}.}

\section{Disclosure Statement}

No conflict of interest exists in this manuscript, and the manuscript was approved by all authors for publication.

\rv{The authors used Claude (Anthropic, Claude Opus 5) to polish the language of the manuscript. The tool was used only for wording and readability; all research design, analysis, results and conclusions are the authors' own, and the authors take full responsibility for the content of the manuscript. The use of large language models as the object of study is described in Section~\ref{subsec:sim_design} and \ref{appendix:geoevolve_config}.}

\appendix
\renewcommand{\thefigure}{S\arabic{figure}}
\setcounter{figure}{0}
\setcounter{table}{0}

\section{Details of Real-world Datasets}\label{appendix:details_datasets}
In this section, we present details on the real-world datasets, including geographical locations, regional divisions, target variables, and associated explanatory variables. In the health dataset, we use eight conditions as target variables: cancer, arthritis, high blood pressure, asthma, depression, diabetes, obesity, and stroke.

\begin{table}[H]
 \scriptsize
 \centering
 \begin{tabular}{ccc}
 \toprule
 \textbf{Column Name} & \textbf{Real Name} & \textbf{Description} \\
 \midrule
 lat & gauge\_lat & Latitude \\
 lon & gauge\_lon & Longitude \\
 region & state & USA state name \\
 Y & t2m & Temperature of air at 2m above the surface \\
 X1 & tp & Total precipitation \\
 X2 & sp & Surface pressure \\
 X3 & u10 & 10m u-component of wind \\
 X4 & v10 & 10m v-component of wind \\
 X5 & pev & Potential evaporation \\
 X6 & strd & Surface thermal radiation downwards \\
 X7 & ssrd & Surface solar radiation downwards \\
 X8 & lai\_hv & Leaf area index, high vegetation \\
 X9 & lai\_lv & Leaf area index, low vegetation \\
 \bottomrule
 \end{tabular}
 \caption{Details of the climate dataset}
 \label{tab:details_climate}
\end{table}

\begin{table}[H]
 \scriptsize
 \centering
 \begin{tabular}{ccc}
 \toprule
 \textbf{Column Name} & \textbf{Real Name} & \textbf{Description} \\
 \midrule
 lat & gauge\_lat & Latitude \\
 lon & gauge\_lon & Longitude \\
 region & region & 2-digit USGS hydrologic unit code \\
 Y & q\_mean & Mean daily discharge \\
 X1 & p\_mean & Mean daily precipitation \\
 X2 & pet\_mean & Mean daily potential evapotranspiration \\
 X3 & aridity & Aridity index (PET/P), the ratio of mean PET to mean precipitation \\
 X4 & elev\_mean & catchment mean elevation \\
 X5 & slope\_mean & catchment mean slope \\
 X6 & frac\_forest & forest fraction \\
 X7 & lai\_max & maximum monthly mean of the leaf area index \\
 X8 & soil\_depth\_statsgo & soil depth (maximum 1.5m, layers marked as water and bedrock were excluded) \\
 X9 & sand\_frac & sand fraction \\
 X10 & soil\_porosity & volumetric porosity \\
 \bottomrule
 \end{tabular}
 \caption{Details of the hydrology dataset}
 \label{tab:details_hydro}
\end{table}

\begin{table}[H]
 \scriptsize
 \centering
 \begin{tabular}{ccc}
 \toprule
 \textbf{Column Name} & \textbf{Real Name} & \textbf{Description} \\
 \midrule
 proj\_X & proj\_X & X on the projected coordinate system \\
 proj\_Y & proj\_Y & Y on the projected coordinate system \\
 region & state & US States 2-letter code \\
 Y & new\_pct\_dem & The percentage of people who voted for the Democratic Party \\
 X1 & sex\_ratio & sex ratio \\
 X2 & pct\_black & The percentage of African Americans, \\
 X3 & pct\_hisp & The percentage of Hispanics \\
 X4 & pct\_bach & The percentage of the population with a bachelor's degree \\
 X5 & median\_income & Median household income \\
 X6 & pct\_65\_over & The percentage of the population aged sixty-five and over \\
 X7 & pct\_age\_18\_29 & The percentage of the population aged eighteen to twenty-nine \\
 X8 & gini & Gini index \\
 X9 & pct\_manuf & The percentage of the population employed in the manufacturing industry \\
 X10 & ln\_pop\_den & Natural logarithm of population density \\
 X11 & pct\_3rd\_party & The percentage of third-party vote \\
 X12 & turn\_out & The voter turnout \\
 X13 & pct\_fb & The percentage of foreign-born \\
 X14 & pct\_uninsured & The percentage of the health-insured population \\
 \bottomrule
 \end{tabular}
 \caption{Details of the politics dataset}
 \label{tab:details_politics}
\end{table}

\begin{table}[H]
 \scriptsize
 \centering
 \begin{tabular}{ccc}
 \toprule
 \textbf{Column Name} & \textbf{Real Name} & \textbf{Description} \\
 \midrule
 lat & lat & Latitude \\
 lon & lon & Longitude \\
 region & StateAbbr & USA State abbreviation \\
 \hline
 \multirow{8}{*}{Y} & CANCER\_AdjPrev & Prevalence of cancer (non-skin) or melanoma among adults, 2022 \\
 & ARTHRITIS\_AdjPrev & Prevalence of arthritis among adults, 2022 \\
 & BPHIGH\_AdjPrev & Prevalence of high blood pressure among adults, 2022 \\
 & CASTHMA\_AdjPrev & Prevalence of current asthma among adults, 2022 \\
 & DEPRESSION\_AdjPrev & Prevalence of depression among adults, 2022 \\
 & DIABETES\_AdjPrev & Prevalence of diagnosed diabetes among adults, 2022 \\
 & OBESITY\_AdjPrev & Prevalence of obesity among adults, 2022 \\
 & STROKE\_AdjPrev & Prevalence of stroke among adults, 2022 \\
 \hline
 X1 & ACCESS2\_AdjPrev & Prevalence of lack of health insurance among adults aged 18-64 years, 2022 \\
 X2 & CHECKUP\_AdjPrev & Prevalence of routine checkup within the past year among adults, 2022 \\
 X3 & DENTAL\_AdjPrev & Prevalence of visited dentist or dental clinic in the past year among adults, 2022 \\
 X4 & BINGE\_AdjPrev & Prevalence of binge drinking among adults, 2022 \\
 X5 & CSMOKING\_AdjPrev & Prevalence of current cigarette smoking among adults, 2022 \\
 X6 & LPA\_AdjPrev & Prevalence of no leisure-time physical activity among adults, 2022 \\
 X7 & SLEEP\_AdjPrev & Prevalence of short sleep duration among adults, 2022 \\
 X8 & EMOTIONSPT\_AdjPrev & Prevalence of lack of social and emotional support among adults, 2022 \\
 \bottomrule
 \end{tabular}
 \caption{Details of the health dataset}
 \label{tab:details_health}
\end{table}

\section{Configuration of GeoEvolve Used in the Case Study}\label{appendix:geoevolve_config}
{
This appendix records the runtime configuration used to evolve the spatial regime algorithms reported in this paper, so that the case study can be reproduced without consulting \citet{luo2025geoevolve}. Settings not listed here are inherited from that system paper without modification.

\subsection*{B.1 LLM backbone and evolutionary search}
The framework is LLM-agnostic in principle; any chat-completions backbone may be substituted via a single configuration entry. In this case study we used a weighted pair of OpenAI models accessed through OpenRouter: \texttt{openai/gpt-4o} as the primary model (weight 0.8) and \texttt{openai/gpt-4.1} as the secondary model (weight 0.2), with sampling parameters \texttt{temperature}\,=\,0.7, \texttt{top\_p}\,=\,0.95, and \texttt{max\_tokens}\,=\,8192. The same backbone is used by all agents (Code Evolver, Code Analyzer, Knowledge Retriever, and Geo-informed Prompt Generator). The code evolver is OpenEvolve \citep{openevolve}, which serves as the inner loop. \nv{Concretely, all GeoEvolve-based variants are run through the open-source Python package \texttt{geoevolve}, distributed alongside \citet{luo2025geoevolve}; the package handles the orchestration of the four agents and exposes the configuration entries described below.} The total evolutionary budget is held constant at 100 iterations across all variants: for OpenEvolve-based variants this is a single flat loop of 100 iterations; for GeoEvolve-based variants it is 10 outer-loop cycles, each invoking the analyzer, retriever (when enabled), and prompt generator once, with 10 inner-loop iterations per cycle. The search uses a MAP-Elites population of 50 programs with archive size 20, organized into 5 islands, with elite/exploration/exploitation selection ratios of $(0.1, 0.2, 0.7)$ and a $10\times10$ \emph{(score, complexity)} feature map. Each evaluation is allowed up to 100 seconds and at most three retries.

\subsection*{B.2 Algorithm variants}
Five variants are compared in addition to the unevolved seed: (i) \textbf{OpenEvolve (No GeoKnow.)}, the vanilla code-evolution loop with no domain-knowledge injection; (ii) \textbf{OpenEvolve (General GeoKnow.)}, a fixed natural-language hint about general spatial-modeling concerns (Tobler's first law, spatial autocorrelation, scale effects); (iii) \textbf{OpenEvolve (Specific GeoKnow.)}, a fixed hint additionally listing spatial-regime-specific modifications such as contiguity penalties and region-size balancing; (iv) \textbf{GeoEvolve (No GeoKnowRAG)}, the full multi-agent dialogue with the retrieval module disabled; and (v) \textbf{GeoEvolve}, the complete pipeline with GeoKnowRAG enabled. The prompt templates for (ii) and (iii) match those in \citet[Section~4]{luo2025geoevolve}.

\subsection*{B.3 GeoKnowRAG and code analyzer}
The geospatial knowledge database is reused unchanged from \citet{luo2025geoevolve}: \rv{124 documents (75{,}228 chunks in the vector store, verified against the archived database used for the reported runs)} drawn from peer-reviewed publications, Wikipedia entries, and open-source GitHub repositories, organized into five categories (geostatistics, spatial theory, GIScience, spatial statistics, and spatial modeling). Representative keywords span kriging theory, Tobler's first law, spatial autocorrelation, the modifiable areal unit problem, and spatial-interaction models. \rv{Chunks of 300 tokens (50-token overlap) are embedded with \texttt{text-embedding-3-large}}, stored in a Chroma vector database, and retrieved with RAG-Fusion (reciprocal rank fusion over multi-angle sub-queries). The Code Analyzer is prompted to (i) identify missing or problematic geographic knowledge, (ii) suggest \rv{up to three} search queries for the retriever, and (iii) flag whether code examples or theoretical background are needed; it returns strict JSON with keys \texttt{missing\_or\_problematic\_knowledge}, \texttt{search\_queries}, \texttt{need\_code\_examples}, and \texttt{need\_geographical\_theory}.

\subsection*{B.4 Task-specific settings for spatial regime discovery}
The settings below differ from \citet{luo2025geoevolve} only because the case study addresses a different task. The \textbf{seed algorithm} $\mathcal{A}_0$ is Two-stage K-Models \citep{guo2023extending} (Section~\ref{subsec:problem_def}). The \textbf{simulator} produces 150 grids of size $25 \times 25$ (Section~\ref{subsec:sim_design}), evenly split across three latent regime patterns (Rectangular, Voronoi, Arbitrary) at the high-noise level. The \textbf{fitness signal} is the Rand Index against the ground-truth partition (Section~\ref{subsec:sim_design}); no SSR-family metric is used as a fitness signal. The \textbf{target number of regions} is $p = 5$ on simulated data, and on real-world data $p$ is inferred from the dataset's administrative-unit structure (Section~\ref{subsec:real_design}). The \textbf{spatial weights matrix} is a rook-contiguity matrix from \texttt{libpysal.weights.lat2W} on the $25\times25$ grid for simulated data, and a row-standardized $k$-nearest-neighbor matrix with $k=6$ for real-world data. The \textbf{minimum region size} is 10 cells (simulated) and 10 observations (real-world). The \textbf{evaluation metrics} (SSR, SSR Reduction, RMSE, Variance Ratio) are computed post-hoc by the evaluation harness and are never observed by $\Phi$ during evolution; \rv{on real-world data they are model-fit statistics rather than predictive scores, for the reasons set out in Section~\ref{subsec:real_design}}.
}

\rv{\section{Computational Cost of the Case Study}\label{appendix:cost}
This appendix reports what it cost to run the case study, separating two quantities that are easily conflated: the cost of \emph{discovering} an algorithm (the evolution loop) and the cost of \emph{running} the discovered algorithm (inference). Both are measured, not estimated.

Token counts are recorded per API call by instrumenting the client SDK, so they cover the inner evolution loop, the four GeoEvolve agents, and GeoKnowRAG retrieval and embedding calls alike. We report token counts rather than monetary cost because tokens are a property of the run, whereas prices depend on the provider and change over time; a reader can convert to their own provider's rates. (At the provider list prices in effect in August 2026, the measured usage corresponds to roughly US\$6--11 per 100-iteration run: US\$5.5--6.4 for the GeoEvolve variants and US\$10.3--11.3 for the OpenEvolve variants.) \rv{\begin{table}[ht]
 \scriptsize
 \centering
 \begin{tabular}{@{}l rrr r r@{}}
 \toprule
 \multirow{3}{*}{\textbf{Variant}} & \multicolumn{4}{c}{\textbf{Evolution} (one 100-iteration run)} & \textbf{Inference} \\
 \cmidrule(lr){2-5}\cmidrule(lr){6-6}
 & \multirow{2}{*}{\textbf{API calls}} & \multicolumn{2}{c}{\rw{\textbf{Tokens (millions)}}} & \multirow{2}{*}{\rw{\textbf{Elapsed time (min)}}} & \multirow{2}{*}{\rw{\textbf{Seconds per grid}}} \\
 \cmidrule(lr){3-4}
 & & \rw{\textbf{Input}} & \rw{\textbf{Output}} & & \\
 \midrule
 GeoEvolve (GeoKnowRAG) & 213 & 1.66 & 0.17 & 32 & \rw{$69.95 \pm 103.05$ ($10.45^{\dagger}$)} \\
 GeoEvolve (No GeoKnowRAG) & 100 & 1.88 & 0.19 & 52 & 1.68 \\
 OpenEvolve (General GeoKnow.) & 100 & 3.82 & 0.18 & \rw{20} & 0.91 \\
 OpenEvolve (Specific GeoKnow.) & 100 & 3.97 & 0.18 & \rw{11} & 7.42 \\
 OpenEvolve (No GeoKnow.) & 100 & 3.69 & 0.17 & \rw{10} & 2.99 \\
 \bottomrule
 \end{tabular}
 \caption{Average measured computational cost per variant. \emph{Evolution} columns refer to one 100-iteration evolution run; \emph{Inference} is the runtime of the discovered algorithm on one 625-cell grid. \rw{For GeoEvolve (GeoKnowRAG) the inference column reports the mean $\pm$ SD over all three evolution runs, followed in parentheses by the trimmed mean$^{\dagger}$ that excludes the one run whose refinement step lacks early stopping (tail behaviour described in the text); no other variant has any run excluded from any column.} Measurement conditions are described in the text.}
 \label{tab:cost}
\end{table}}

\rw{Elapsed time is measured from the start of an evolution run to the completion of its final iteration, and inference time is the mean runtime of a discovered algorithm per grid over the 150 grids. All timings were measured on one machine (Apple M5); inference used six worker processes.}

Three observations follow from Table~\ref{tab:cost}. First, a complete 100-iteration evolution is inexpensive in absolute terms: under $4$M tokens and roughly an hour of \rw{elapsed time} at most for every variant. Discovery cost is therefore not the binding constraint on this kind of work; run-to-run variance is (Section~\ref{subsec:sim_results}).

Second, the GeoEvolve variants consume roughly half the input tokens of the OpenEvolve variants ($1.7$--$1.9$M versus $3.7$--$4.0$M) despite issuing about twice as many API calls. The reason is structural rather than a property of retrieval: GeoEvolve restarts the inner loop every ten iterations, so the program database --- from which the prompt draws its top-performing and diverse exemplars --- is rebuilt each round and supplies fewer exemplars in the early iterations of every round. A flat 100-iteration loop instead accumulates a full archive and quickly saturates the prompt. Measured per-call prompt length shows this directly: GeoEvolve prompts follow a sawtooth from ${\sim}7$k tokens at the start of each round to ${\sim}40$k by its end (median ${\sim}22$k), whereas OpenEvolve prompts sit near ${\sim}42$k throughout. Disabling GeoKnowRAG leaves this profile essentially unchanged (median $21.7$k versus $21.8$k with retrieval enabled), confirming that the saving comes from the two-loop schedule and not from retrieval compression.

Third, inference cost varies far more across variants than discovery cost does, and the ordering does not follow accuracy. The overhead of the GeoEvolve algorithm relative to the seed comes from two explicit and adjustable choices --- the number of candidate cluster counts searched in stage one, and the iteration budget and patience of the border-refinement step --- rather than from any fixed property of the framework. \rw{One evolved program illustrates the risk of leaving those choices unconstrained: its refinement step has no early stopping, so although its median grid takes $4.8$~s, it exceeds ten minutes on 19 of the 150 grids, for a per-grid mean of $188.9$~s. The missing early stopping makes it slow, not unreliable: like every other frozen program, it completes all 150 grids. This slow behaviour is itself an outcome of the discovery process, so Table~\ref{tab:cost} reports the inference cost of GeoEvolve (GeoKnowRAG) over all three runs, with the clearly marked trimmed mean that excludes this run summarizing typical behaviour rather than the tail.} Evolution optimizes the fitness signal it is given, and runtime was not part of that signal; a multi-objective fitness that prices runtime would be a natural extension (Section~\ref{sec:discussion}).}

\rw{\section{Evolved Algorithm versus Seed under a Matched Budget, and Component Ablations}\label{appendix:equalbudget}
Compared with its seed, the evolved algorithm spends more computation at inference: it searches six initial cluster counts and adds a border-refinement stage. This appendix asks whether its advantage in Table~\ref{tab:baseline_simulation} comes from that extra computation or from its algorithmic content. Two analyses answer the question: a comparison with the seed given the same search under a matched budget, and one-component ablations of the evolved algorithm.

\textbf{ARI, not SSR, is the comparison of record in this appendix.} The reason is structural. Because no ground-truth information may enter a candidate program (Principle 4, Section~\ref{subsec:module3}), every selection rule available in this study --- inside the evolved algorithm and inside the matched-budget wrapper alike --- must rank candidate partitions by post-merge SSR. SSR is therefore the one quantity that additional search can buy directly, and a search-based baseline will narrow the SSR gap whether or not it has recovered any more of the regime structure. ARI, agreement with the ground-truth partition, is reachable by no selection rule here, and it is the quantity spatial regime discovery is defined on. Closeness on SSR between a search-based baseline and the evolved algorithm is accordingly weak evidence of equivalence, and the two columns are read separately below.

All variants are evaluated by the harness of Section~\ref{subsec:sim_design} on the same 150 grids, in a single batch on one machine. The reference row is the single frozen algorithm $\mathcal{A}^\star$ that is carried forward to the real-world study, scored grid by grid; its figures are therefore not the three-run means of Table~\ref{tab:baseline_simulation}, where the spread is across evolution runs rather than across grids. Table~\ref{tab:equalbudget} reports the results; the programs, the runner, and the per-grid results are in the project repository.

\paragraph{Evolved algorithm versus seed under a matched budget} This analysis tests whether extra search alone reproduces the evolved algorithm's advantage. The baseline is the seed program, left unmodified, called by an external wrapper. The wrapper repeats the seed's partition-and-merge procedure over the same six initial cluster counts that the evolved algorithm searches, $\{\lfloor 2.5p \rfloor, 4p, \lfloor\sqrt{n}\rfloor, p, p{+}1, p{+}2\}$, draws a fresh random restart on each pass, and returns the run with the lowest post-merge SSR, which is the label-free selection rule the evolved algorithm itself uses. It stops once it has used the evolved algorithm's mean per-grid runtime in the same batch ($7.5$~s). The remaining asymmetries favour the baseline: it always completes at least one full pass over the six counts, and it checks the budget only between runs, so it actually takes $1.11$ times as long as the evolved algorithm. The outcome divides along exactly the line drawn above. On ARI, which no selection rule in this study can target, the gap persists in all three spatial patterns: the evolved algorithm recovers the true regimes better ($0.894$ versus $0.854$; median $\Delta$ARI $+0.043$, $p = 2.3\times10^{-10}$), and the budgeted search recovers only a little over half of the seed's deficit ($0.801 \rightarrow 0.854$, against $0.801 \rightarrow 0.894$ for the evolved algorithm). On SSR, the quantity the wrapper selects on directly, additional search closes most of the gap (seed $80.5$, baseline $60.6$, evolved $59.0$), leaving a small but significant margin (median $\Delta$SSR $-1.90$, $p = 2.1\times10^{-5}$). Additional computation therefore accounts for most of the SSR advantage but not for the recovery of the regime structure itself, and the asymmetry between the two columns is what a baseline that optimizes one of them and not the other should be expected to produce.

\paragraph{Component ablations} The ablations locate the advantage that remains, and they do so in the direction opposite to what a parameter-search account of the gains would predict. Each removes one component from the frozen evolved algorithm with a single commented change: the initial-cluster-count search (replaced by the seed's fixed count of 20); border refinement (the stage that reassigns boundary units after merging); the spatial penalty inside that stage (which discourages moving a unit into a region where it has fewer neighbours); or KMeans++ initialization (replaced by the seed's random assignment).

The one component that is purely a matter of search effort, the initial-cluster-count search, contributes the least of the three that contribute at all (median $\Delta$SSR $-0.33$, $\Delta$ARI $+0.004$), while the structural addition contributes an order of magnitude more (border refinement: median $\Delta$SSR $-7.09$, $\Delta$ARI $+0.063$, $p < 10^{-24}$ on both). KMeans++ initialization contributes nothing detectable ($p = 0.68$) while accounting for $43\%$ of the runtime --- a component the search had no reason to prune, because runtime was not part of the fitness signal (\ref{appendix:cost}). The spatial penalty is the clearest case of all: it contributes $+0.039$ ARI even though removing it \emph{lowers} SSR (median $\Delta$SSR $+1.80$, $p = 3.9\times10^{-19}$), trading fit for agreement with the true structure. Because its value is negative in the algorithm's own selection objective, no tuning of the seed's parameters against that objective could have found it; it is a structural contribution of the discovery process rather than a setting a grid search would have reached.

\begin{table}[ht]
 \scriptsize
 \centering
 \setlength{\tabcolsep}{3.5pt}
 \begin{tabular}{@{}l cc r rr rr@{}}
 \toprule
 \multirow{2}{*}{\textbf{Variant}} & \multirow{2}{*}{\textbf{SSR} $\downarrow$} & \multirow{2}{*}{\textbf{ARI} $\uparrow$} & \multirow{2}{*}{\textbf{Rel.\ time}} & \multicolumn{2}{c}{\textbf{$\Delta$SSR}} & \multicolumn{2}{c}{\textbf{$\Delta$ARI}} \\
 \cmidrule(lr){5-6}\cmidrule(lr){7-8}
 & & & & median & $p$ & median & $p$ \\
 \midrule
 Evolved algorithm (reference)          & $58.96 \pm 9.46$  & $0.894 \pm 0.096$ & $1.00$ & & & & \\
 \midrule
 \multicolumn{8}{@{}l}{\emph{(a) Seed under a matched budget}} \\
 \quad Seed, single run                 & $80.52 \pm 20.11$ & $0.801 \pm 0.105$ & $0.10$ & $-17.87$ & $1.1\times10^{-24}$ & $+0.084$ & $6.0\times10^{-21}$ \\
 \quad Seed + matched-budget search     & $60.60 \pm 8.64$  & $0.854 \pm 0.091$ & $1.11$ & $-1.90$  & $2.1\times10^{-5}$  & $+0.043$ & $2.3\times10^{-10}$ \\
 \midrule
 \multicolumn{8}{@{}l}{\emph{(b) Evolved algorithm with one component removed}} \\
 \quad w/o initial-$k$ search           & $62.35 \pm 12.08$ & $0.873 \pm 0.114$ & $0.56$ & $-0.33$  & $4.3\times10^{-5}$  & $+0.004$ & $6.3\times10^{-3}$ \\
 \quad w/o border refinement            & $68.72 \pm 13.68$ & $0.830 \pm 0.094$ & $0.98$ & $-7.09$  & $2.1\times10^{-25}$ & $+0.063$ & $2.4\times10^{-26}$ \\
 \quad w/o spatial penalty              & $56.54 \pm 8.53$  & $0.852 \pm 0.092$ & $0.99$ & $+1.80$  & $3.9\times10^{-19}$ & $+0.039$ & $1.1\times10^{-24}$ \\
 \quad w/o KMeans++ initialization      & $57.90 \pm 7.09$  & $0.902 \pm 0.092$ & $0.57$ & $0.00$   & $0.68$              & $0.000$  & $0.68$ \\
 \bottomrule
 \end{tabular}
 \caption{Matched-budget comparison and one-component ablations on the 150 high-noise grids. \textbf{ARI is the comparison of record: SSR is the quantity every selection rule in this study optimizes directly (Principle 4), so closeness on SSR does not indicate equivalence.} SSR and ARI are mean $\pm$ SD over grids, computed externally by the harness. $\Delta$SSR and $\Delta$ARI are per-grid medians of (evolved $-$ variant), each with the $p$-value of a paired Wilcoxon signed-rank test; negative $\Delta$SSR and positive $\Delta$ARI favour the evolved algorithm. Paired medians are reported because the per-grid SSR distribution is right-skewed, so they need not match the difference of the means in the first column. Rel.\ time is each program's mean per-grid runtime relative to the evolved algorithm ($=1.00$), all measured in the same batch. The two identical $p$-values in the last row arise from the signed-rank statistic taking mirror-image values on the two metrics; they are a coincidence, not a transcription error.}
 \label{tab:equalbudget}
\end{table}}

\section{Worked Example of the Automatic Code Formatter on Two-stage K-Models}\label{appendix:formatter_example}
\begin{figure}[H]
 \centering
 \includegraphics[width=0.65\linewidth]{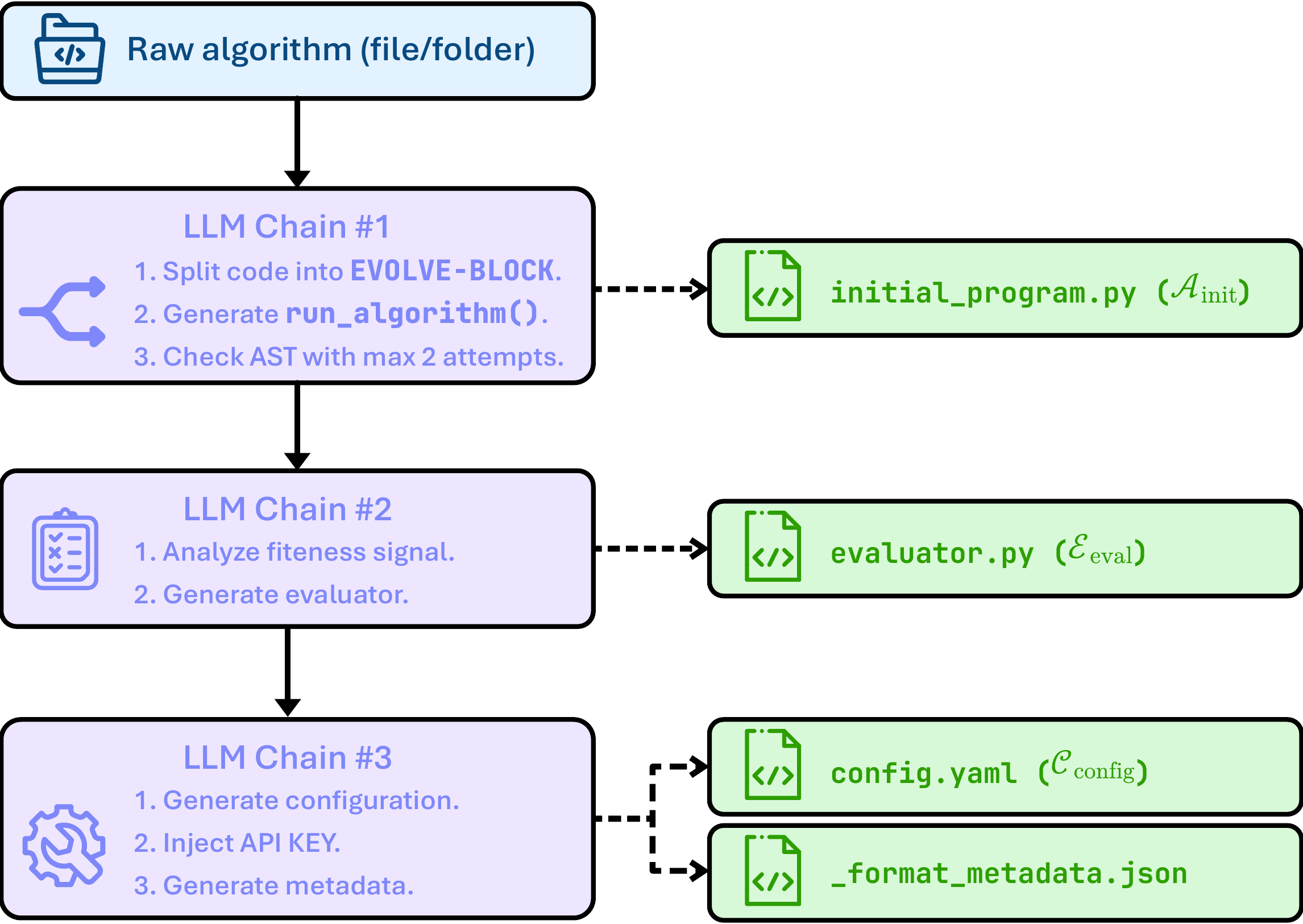}
 \caption{Trace of the automatic code formatter on the Two-stage K-Models seed. Three LLM chains transform the raw \texttt{.py} input into the standardized triple plus a metadata file: \texttt{initial\_program.py} ($\mathcal{A}_{\text{init}}$), \texttt{evaluator.py} ($\mathcal{E}_{\text{eval}}$), \texttt{config.yaml} ($\mathcal{C}_{\text{config}}$), and \texttt{\_format\_metadata.json}. Chain~\#1 is followed by a deterministic syntactic check that parses the generated program with Python's \texttt{ast} module, locates the standardized \texttt{run\_algorithm} entry function, and verifies that its body contains at least one \texttt{return} statement returning a non-\texttt{None} value; when the check fails, an LLM-driven repair step is invoked at most twice before the formatter falls back to the user. For this seed the check passed on the first attempt.}
 \label{fig:formatter_pipeline}
\end{figure}

This appendix traces what the automatic code formatter (Module~\modTwo, Section~\ref{subsec:module2}) did when applied to the Two-stage K-Models seed. The formatter's job is to turn the raw source into the standardized triple $(\mathcal{A}_{\text{init}}, \mathcal{E}_{\text{eval}}, \mathcal{C}_{\text{config}})$ introduced in Section~\ref{subsec:module2}, plus a small metadata file. Figure~\ref{fig:formatter_pipeline} summarizes the end-to-end run; Sections~\ref{subsec:formatter_example_input}--\ref{subsec:formatter_example_narrative} below show the raw input, the three produced artifacts, and a short narrative of the case-study-specific elements.

\subsection{Input: the raw Two-stage K-Models implementation}\label{subsec:formatter_example_input}
The raw input $\mathcal{A}_{\text{raw}}$ was a single \texttt{.py} file containing our implementation of the Two-stage K-Models algorithm of \citet{guo2023extending}. The core of the file is the two-stage entry point, reproduced in abbreviated form in Listing~\ref{lst:raw_2kmodels}: a first-stage \texttt{run\_kmodels} call that partitions the units into 20 sub-regions, followed by a second-stage \texttt{merge\_stage} that enforces spatial contiguity, removes regions below \texttt{min\_size}, and merges until exactly $p$ regions remain. The file also contains the underlying helper functions (\texttt{init\_zones}, \texttt{fit\_equations}, \texttt{closest\_equation}, \texttt{merge\_small\_regions}, \texttt{greedy\_merge\_until\_p}, \dots) and the data-loading / metric-computation utilities. Crucially, the raw file does not declare which parts of the code should be considered mutable by an evolutionary loop and does not expose a uniform entry point for an external evaluator, those are the two responsibilities the formatter takes on.

\begin{lstlisting}[language=Python, caption={Excerpt from the raw Two-stage K-Models source $\mathcal{A}_{\text{raw}}$ given to the formatter (helper functions and data-loading utilities elided).}, label={lst:raw_2kmodels}]
import numpy as np
import libpysal

def run_two_stage_kmeans(
 X, Y, p, w,
 min_size=None, max_iter=1000,
 init_stoc_step=True, verbose=False):
 """A two-step kmeans-like regionalization algorithm."""
 if min_size is None:
 min_size = X.shape[1]
 # Stage 1: partition into 20 sub-regions
 labels_partition = run_kmodels(
 X, Y, n_regions=20, w=w,
 min_size=min_size, max_iter=max_iter,
 init_stoc_step=init_stoc_step, verbose=verbose)
 # Stage 2: merge into p connected regions
 labels_merged = merge_stage(
 X, Y, labels_partition, w,
 p=p, min_size=min_size, verbose=verbose)
 return labels_merged
\end{lstlisting}

\subsection{Output: the standardized triple}\label{subsec:formatter_example_output}

\subsubsection{Initial program $\mathcal{A}_{\text{init}}$}
The formatter wrapped the algorithmic core of $\mathcal{A}_{\text{raw}}$ in a single \texttt{\# EVOLVE-BLOCK-START} / \texttt{\# EVOLVE-BLOCK-END} pair and appended a standardized entry function named \texttt{run\_algorithm} below the block. The function name is fixed by the formatter (it is the same name the formatter uses for every seed, and the same name the AST-based validation step checks for), so that the evaluator can always import the program and call \texttt{run\_algorithm()} without seed-specific code. The block contains the entire two-stage algorithm together with all helper routines that depend on the regionalization logic; the entry function lives outside the block so that the inner loop cannot mutate the interface that the evaluator relies on. Listing~\ref{lst:Ainit} shows the boundaries of the resulting \texttt{initial\_program.py}.

\begin{lstlisting}[language=Python, caption={Boundaries of the formatter-produced \texttt{initial\_program.py} ($\mathcal{A}_{\text{init}}$). The evolvable block contains the partitioning loop, the merge stage, and all helper functions; the standardized \texttt{run\_algorithm()} entry function below the block is what the evaluator calls.}, label={lst:Ainit}]
# EVOLVE-BLOCK-START
import time
import numpy as np
import libpysal

def run_two_stage_kmeans(X, Y, p, w, min_size=None,
 max_iter=1000, init_stoc_step=True,
 verbose=False):
 ... # algorithmic core
 return labels_merged

def run_kmodels(...): ... # Stage 1 (partition)
def merge_stage(...): ... # Stage 2 (merge)
def init_zones(...): ... # initialization
def fit_equations(...): ... # per-region OLS
# ... other helpers ...
# EVOLVE-BLOCK-END

# --- standardized entry function (not mutated by the inner loop) ---
def run_algorithm():
 """Run Two-stage K-Models on every simulated grid and return
 the mean of the five evaluator-facing metrics."""
 base_path = '.../synthetic'
 data_ids = (5, 75, 149)
 ssr_list, randi_list, nmi_list, mae_list, time_list = [],[],[],[],[]
 for id_ in data_ids:
 data_path = f'{base_path}/grid_{id_}h.txt'
 ssr, randi, nmi, mae, used_time = run_georegime_once(data_path)
 ssr_list.append(ssr); randi_list.append(randi)
 nmi_list.append(nmi); mae_list.append(mae)
 time_list.append(used_time)
 return (np.mean(ssr_list), np.mean(randi_list),
 np.mean(nmi_list), np.mean(mae_list),
 np.mean(time_list))
\end{lstlisting}

\subsubsection{Evaluator module $\mathcal{E}_{\text{eval}}$}
The formatter wrote a \texttt{evaluator.py} module whose \texttt{evaluate(program\_path)} entry point dynamically loads $\mathcal{A}_{\text{init}}$ with \texttt{importlib}, calls \texttt{program.run\_algorithm()} on the simulator under a 100-second wall-clock timeout, and returns a dictionary of five metrics together with a \texttt{combined\_score} field that is the single scalar actually consumed by the inner loop. Listing~\ref{lst:Eeval} shows the body of \texttt{evaluate()}. The key design choice visible in the listing is that \texttt{combined\_score} is set to \texttt{randi}: the Rand Index is what drives evolutionary selection, while SSR, NMI, MAE, and \texttt{used\_time} are returned and logged but never fed back as fitness. This is the code-level enforcement of the fitness/evaluation separation prescribed by the training protocol of Section~\ref{subsec:module3}.

\begin{lstlisting}[language=Python, caption={Body of the formatter-produced \texttt{evaluator.py} ($\mathcal{E}_{\text{eval}}$). The \texttt{combined\_score} field (last line of the returned dict) is what the inner-loop engine uses for selection; all other metrics are logged but do not influence the search.}, label={lst:Eeval}]
def evaluate(program_path):
 spec = importlib.util.spec_from_file_location("program", program_path)
 program = importlib.util.module_from_spec(spec)
 spec.loader.exec_module(program)
 if not hasattr(program, "run_algorithm"):
 print("Error: program lacks the expected entry function")
 ssr, randi, nmi, mae, used_time = run_with_timeout(
 program.run_algorithm,
 timeout_seconds=100)
 return {"ssr": ssr,
 "randi": randi, # fitness signal (RandI vs ground truth)
 "nmi": nmi,
 "mae": mae,
 "used_time": used_time,
 "combined_score": randi} # drives evolutionary selection
\end{lstlisting}

\subsubsection{Configuration file $\mathcal{C}_{\text{config}}$}
The formatter emitted a YAML configuration file with five blocks: general evolution settings, LLM settings, prompt configuration, population/archive parameters for OpenEvolve's MAP-Elites database, and evaluator settings. The most case-study-specific block is the prompt configuration, where the formatter inserted a domain-specific system message instructing the LLM that the task is to improve a spatial regime algorithm and that the optimization target is the Rand Index. Listing~\ref{lst:Cconfig} shows the relevant blocks (full file: 100 lines). The LLM-generated YAML initially contains the literal placeholder \texttt{"YOUR\_API\_KEY"} for \texttt{llm.api\_key}; the formatter then post-processes the file to replace the placeholder with whichever of \texttt{OPENROUTER\_API\_KEY}, \texttt{OPENAI\_API\_KEY}, \texttt{GEMINI\_API\_KEY}, \texttt{GROQ\_API\_KEY}, or \texttt{ANTHROPIC\_API\_KEY} is present in the environment (resolved in that priority order), and to rewrite \texttt{llm.api\_base} so it matches the backend implied by the model name (e.g.\ OpenRouter, Gemini's OpenAI-compatible endpoint, or Groq). The injection happens at write time so that no secret ever lives in the source tree or in version control.

\begin{lstlisting}[language={[LaTeX]TeX}, caption={Key blocks of the formatter-produced \texttt{config.yaml} ($\mathcal{C}_{\text{config}}$). API keys are written to the file at run time from an environment variable so that no secret enters version control.}, label={lst:Cconfig}]
# --- General evolution settings ---
max_iterations: 10 # inner-loop iterations per outer cycle; GeoEvolve runs 10 cycles for 100 total (Appendix B.1); for OpenEvolve-only variants this is overridden to 100 to run a single flat loop
diff_based_evolution: true
max_code_length: 25000

# --- LLM configuration ---
llm:
 primary_model: "openai/gpt-4o"
 primary_model_weight: 0.8
 secondary_model: "openai/gpt-4.1"
 secondary_model_weight: 0.2
 temperature: 0.7
 top_p: 0.95
 max_tokens: 8192

# --- Prompt configuration (task-specific) ---
prompt:
 system_message: "You are an expert in GIS and an experienced
 coder. Your task is to improve the Spatial Regime algorithm
 in Python ... The target is to maximize Rand index (RandI)."

# --- Database (MAP-Elites + island model) ---
database:
 population_size: 50
 archive_size: 20
 num_islands: 5
 elite_selection_ratio: 0.1
 exploration_ratio: 0.2
 exploitation_ratio: 0.7
 feature_dimensions: ["score", "complexity"]
 feature_bins: 10

# --- Evaluator ---
evaluator:
 timeout: 100
 max_retries: 3
 parallel_evaluations: 4
\end{lstlisting}

\subsubsection{Run metadata \texttt{\_format\_metadata.json}}
Alongside the standardized triple, the formatter writes a small JSON file with the information needed to audit and reverse the formatting run after the fact. For the Two-stage K-Models seed it contains six fields: \texttt{original\_path} (the absolute path of the raw input), \texttt{original\_type} (\texttt{"file"} or \texttt{"directory"}; \texttt{"file"} for this seed), \texttt{original\_files} (a map from each input filename to the SHA-256 hash of its content), \texttt{run\_function\_name} (always \texttt{"run\_algorithm"} in the current formatter), \texttt{description} (the natural-language algorithm description the user supplied when invoking the formatter), and \texttt{created\_at} (an ISO 8601 timestamp). This file is not consumed by the inner-loop engine; it is read by the companion function \texttt{restore\_from\_geoevolve()}, which uses \texttt{original\_path} / \texttt{original\_type} to decide whether to write a single file or a folder and uses \texttt{original\_files} to verify, via hash comparison, that the raw source has not been modified since the formatter was run.

\subsection{Narrative}\label{subsec:formatter_example_narrative}

\paragraph{Evolvable-block identification} The formatter identified a single evolvable region in $\mathcal{A}_{\text{raw}}$ and placed one \texttt{\# EVOLVE-BLOCK-START} / \texttt{\# EVOLVE-BLOCK-END} pair around it. The region it chose corresponds to the partitioning loop, the merge stage, and all of their helper routines, exactly the algorithmic locus that subsequent evolved variants ended up modifying. The data-loading utility (\texttt{input\_data}), the metric-computation class (\texttt{RegionMetrics}), and the per-dataset driver (\texttt{run\_georegime\_once}) were correctly kept outside the block, which is what made it safe for the evolved variants to mutate the algorithm without breaking the evaluator-facing interface.

\paragraph{First-pass correctness of the evaluator} The evaluator generated by the second chain ran successfully on the first attempt: it imported $\mathcal{A}_{\text{init}}$, found the standardized \texttt{run\_algorithm} entry function, called it under the 100-second timeout, and reported the five-metric dictionary together with \texttt{combined\_score} set to \texttt{randi}. The deterministic AST-based validation step on \texttt{run\_algorithm} (Figure~\ref{fig:formatter_pipeline}) was therefore not triggered, and the LLM-driven repair step was not invoked. The reverse operation, \texttt{restore\_from\_geoevolve()}, was not used at this stage of the case study but is available for splicing an evolved \texttt{initial\_program.py} back into the user's original file structure using the hashes and paths recorded in \texttt{\_format\_metadata.json}.

\paragraph{Configuration choices} The configuration file was produced in one pass and reflects the OpenEvolve defaults with one task-specific override: the system message embeds the domain ("Spatial Regime algorithm") and the optimization target ("Rand index"). The MAP-Elites feature dimensions \texttt{score} and \texttt{complexity} were left at their defaults; the LLM choice (GPT-4o as primary with GPT-4.1 as a secondary fallback) and the temperature 0.7 were inherited from the formatter's prompt template and not modified per task.
\color{black}

\section{\nv{Detailed K-Models Variant Comparison}}\label{appendix:realworld_variants}

\begin{figure}[H]
 \centering
 \includegraphics[width=\linewidth]{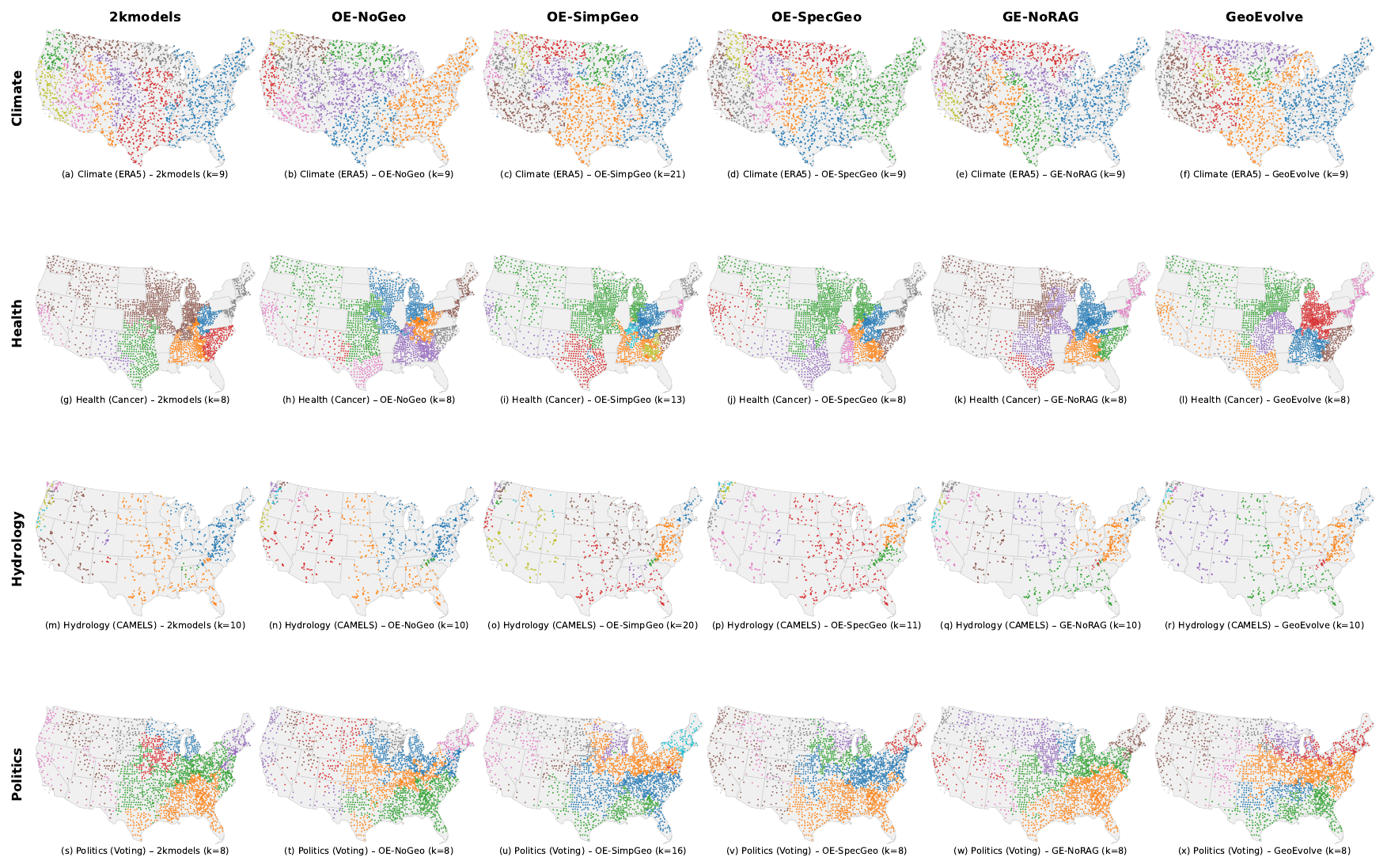}
 \caption{Regime-partitioning maps produced by the six Two-stage K-Models variants on the same four representative real-world datasets, all under the unified evaluation protocol of Section~\ref{subsec:real_design}. Column-header abbreviations are as in Figure~\ref{fig:baseline_comparison_real_world}: \emph{2kmodels} (unevolved seed), \emph{OE-NoGeo} / \emph{OE-SimpGeo} / \emph{OE-SpecGeo} (OpenEvolve variants with no, general, and specific geographic-knowledge prompts), \emph{GE-NoRAG} (GeoEvolve without GeoKnowRAG), and \emph{GeoEvolve} (full GeoEvolve with GeoKnowRAG).}
 \label{fig:evolved_regime_maps}
\end{figure}

\nv{This appendix provides the detailed within-family breakdown of the six Two-stage K-Models variants on real-world data, complementing the main-text comparison of GeoEvolve against the four classical regionalization baselines (Section~\ref{subsec:real_results}) and the per-pattern simulation breakdown already reported in Table~\ref{tab:baseline_simulation}. The real-world breakdown spans all 11 datasets in four application domains: climate, health, hydrology, and politics, and includes a fine-grained six-variant regime-map breakdown (Figure~\ref{fig:evolved_regime_maps}).}

\begin{figure}[H]
 \centering
 \includegraphics[width=\linewidth]{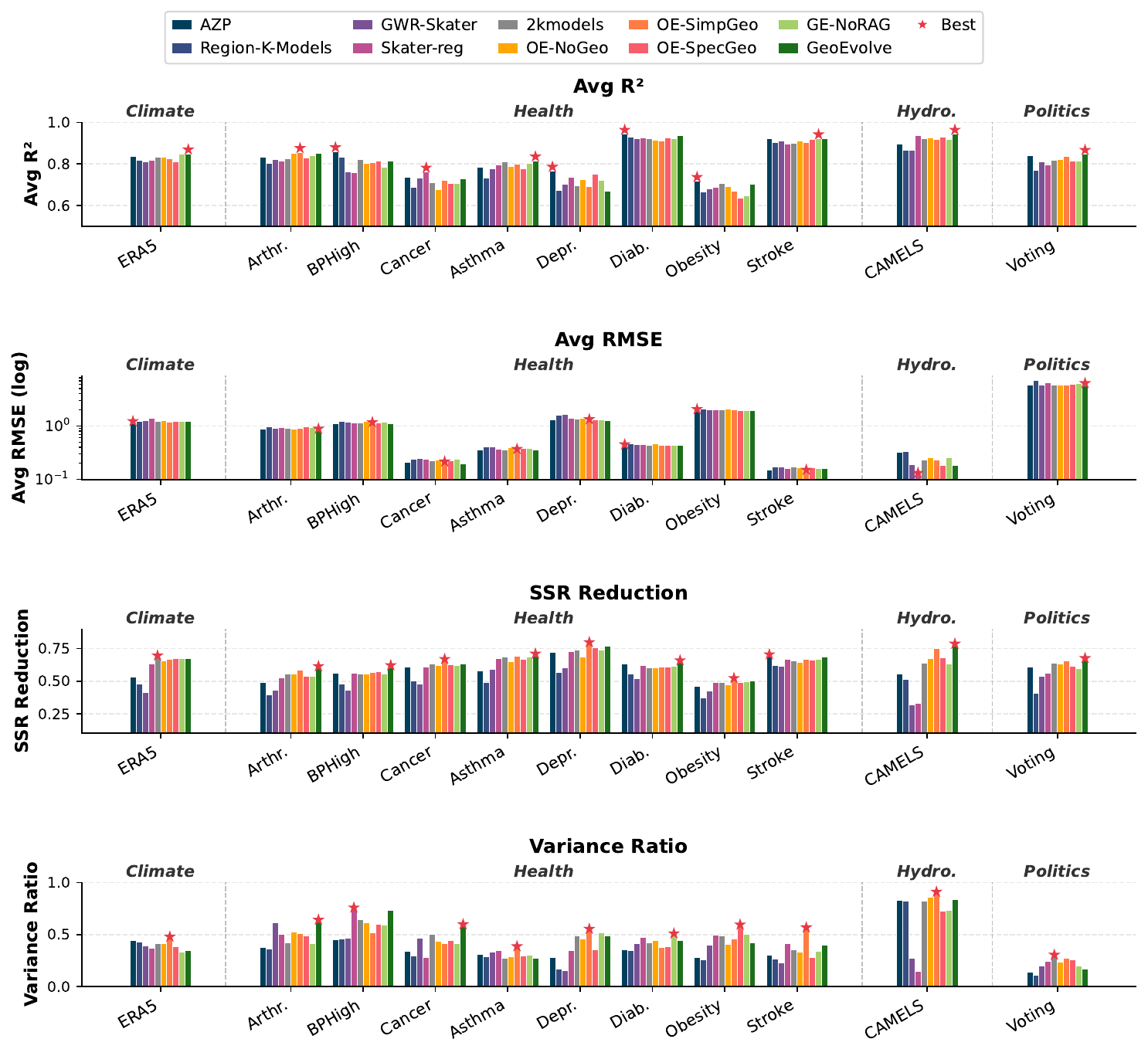}
 \caption{Full ten-algorithm performance comparison across the 11 real-world datasets, grouped by domain. Algorithms are split into four classical regionalization baselines (cool palette) and six Two-stage K-Models variants (warm palette). The legend abbreviations for the K-Models family are: \emph{2kmodels} = unevolved Two-stage K-Models seed; \emph{OE-NoGeo} = OpenEvolve without geographic knowledge; \emph{OE-SimpGeo} = OpenEvolve with general (simple) geographic knowledge prompt; \emph{OE-SpecGeo} = OpenEvolve with task-specific geographic knowledge prompt; \emph{GE-NoRAG} = GeoEvolve without the GeoKnowRAG retrieval module; \emph{GeoEvolve} = the full GeoEvolve pipeline with GeoKnowRAG. Red stars indicate the best algorithm on each dataset and metric.}
 \label{fig:baseline_comparison_real_world_full}
\end{figure}

\nv{\paragraph{Real-world cross-domain bar chart: full ten-algorithm comparison}}
\nv{Figure~\ref{fig:baseline_comparison_real_world_full} expands the compact ten-algorithm SSR-Reduction summary in main-text Figure~\ref{fig:baseline_comparison_real_world} and the six-algorithm Table~\ref{tab:baseline_realworld_summary} into a per-dataset, four-metric breakdown across all 11 datasets, additionally displaying the three OpenEvolve variants and GeoEvolve without GeoKnowRAG (which are absent from Table~\ref{tab:baseline_realworld_summary}) so that readers can see the full within-family detail of the K-Models seed and all its evolved variants alongside the four classical baselines. The patterns visible in this fuller chart are consistent with the main-text summary: GeoEvolve is the most consistent winner on RMSE and SSR Reduction across both the K-Models family and the classical baselines; AZP is the most competitive classical baseline on $R^2$; and the OpenEvolve variant with a general geographic-knowledge prompt attains a high Variance Ratio (mean 0.487, five first-place wins out of 11 datasets), although it does not match GeoEvolve on RMSE or SSR Reduction.}

\begin{figure}[H]
 \centering
 \includegraphics[width=\linewidth]{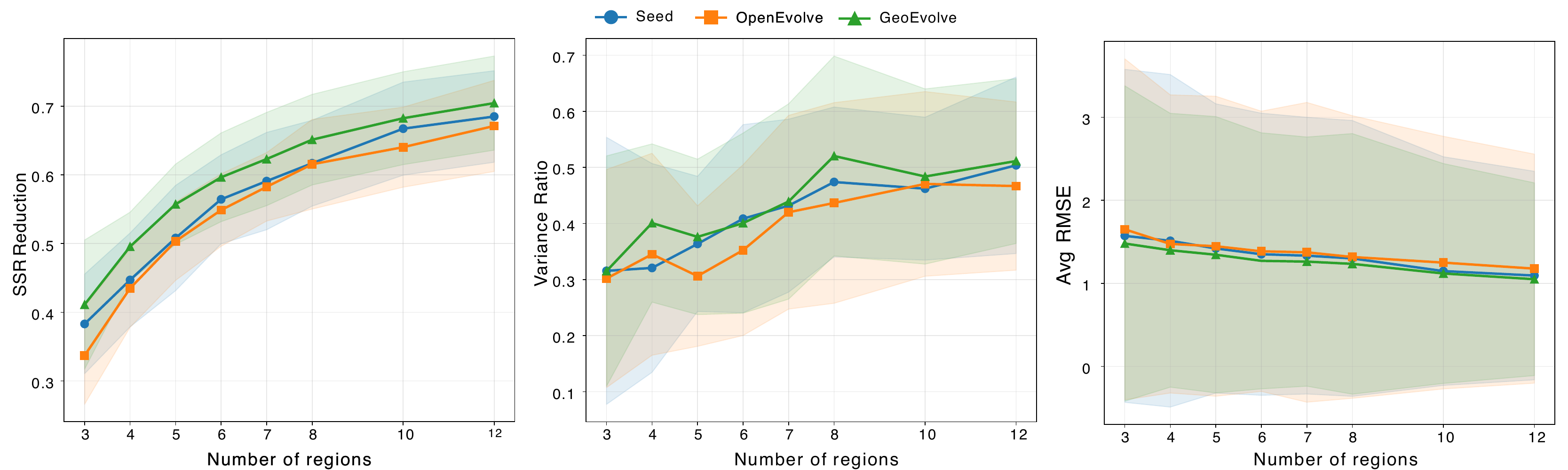}
 \caption{Sensitivity of algorithm performance to the number of spatial regimes. Each panel reports the mean value (solid line) $\pm$ one standard deviation (shaded band) across 11 datasets for SSR Reduction $\uparrow$ (left), Variance Ratio $\uparrow$ (center), and Average RMSE $\downarrow$ (right).}
 \label{fig:p_sensitivity_mean}
\end{figure}

\paragraph{Sensitivity to the number of spatial regimes}
Figure~\ref{fig:p_sensitivity_mean} examines the sensitivity of algorithm performance to the prescribed number of spatial regimes ($R$), ranging from 3 to 12. SSR reduction and the variance ratio increase monotonically across all three algorithms, while average RMSE decreases correspondingly, consistent with the greater flexibility of finer spatial partitions. Across the considered range of $R$, GeoEvolve tends to attain higher SSR reduction and variance ratio, and lower average RMSE. Notably, GeoEvolve at $R=6$ already matches the SSR reduction achieved by the seed 2KModels at $R=8$.

\paragraph{Spatial regime assignments across the six K-Models variants}
\nv{Figure~\ref{fig:evolved_regime_maps} reports the regime maps for all six Two-stage K-Models variants on four representative datasets (Climate / Health / Hydrology / Politics), complementing the classical-baseline maps in main-text Figure~\ref{fig:baseline_regime_maps}. The figure isolates the visual effect of each step along the knowledge-injection design axis (no prompt $\to$ general prompt $\to$ specific prompt $\to$ multi-agent dialogue without retrieval $\to$ full GeoKnowRAG). Several qualitative patterns stand out across the four domains. In the climate domain, all variants produce a similar macro-structure that separates the western arid/semi-arid zone from the more humid east, and GeoEvolve tends to draw sharper boundaries along the Rocky Mountain front. In the health domain (Cancer dataset as an example), the seed algorithm tends to produce a single dominant region spanning the eastern United States, whereas GeoEvolve yields a more balanced partition that distinguishes the Appalachian corridor from the southeastern coastal plain. The hydrology domain shows the greatest agreement across methods, reflecting the strong hydroclimatic zonation of the contiguous United States. For politics, GeoEvolve delineates a sharper urban-corridor versus rural-interior contrast, particularly along the Northeast--Mid-Atlantic megalopolis and the Pacific coast.}

\begin{figure}[H]
 \centering
 \includegraphics[width=\linewidth]{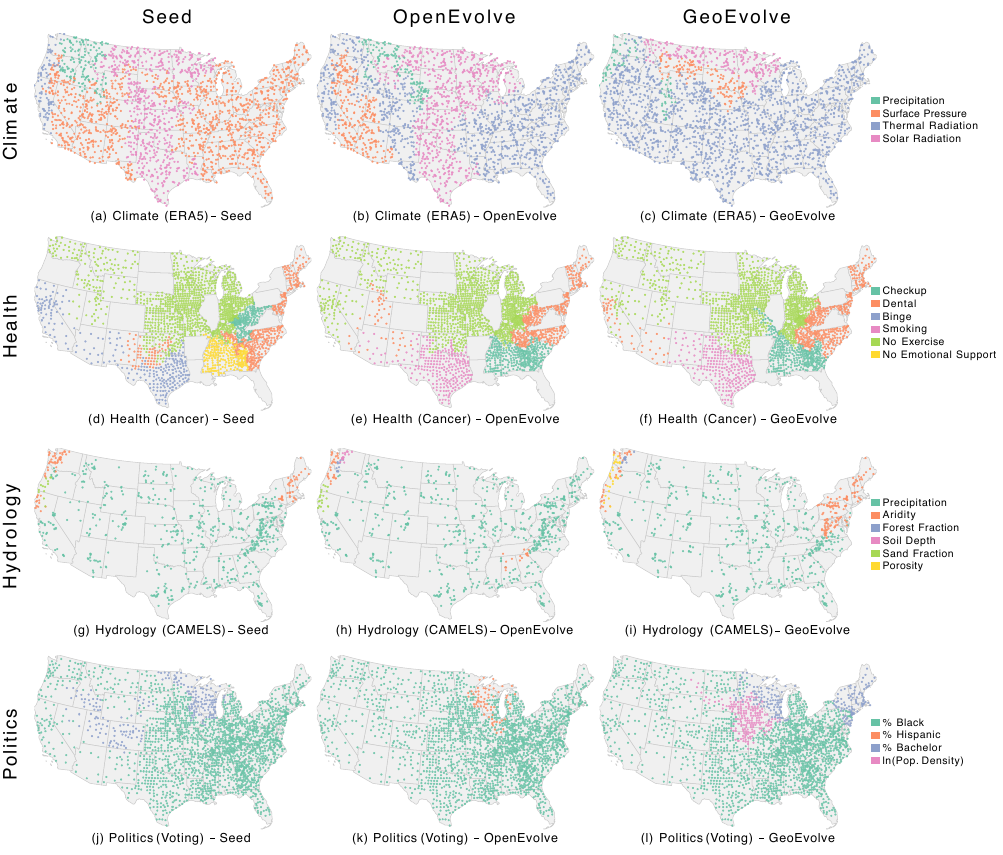}
 \caption{Dominant covariates by spatial regime across datasets and algorithm variants.}
 \label{fig:dominant_covariate_maps}
\end{figure}

\paragraph{Dominant covariate maps}
Figure~\ref{fig:dominant_covariate_maps} maps the dominant covariate, defined as the predictor with the largest absolute standardized coefficient ($\arg\max|\beta|$) excluding the intercept, for each observation. The seed algorithm produces relatively fragmented, locally heterogeneous patterns; OpenEvolve introduces more spatial continuity with partially consolidated regions; GeoEvolve produces more spatially coherent and regionally structured regimes, with clearer clustering and smoother transitions between dominant covariates. For health, hydrology, and politics, the leading drivers across the three algorithms are similar (no exercise, precipitation, and the percentage of Black residents, respectively); in the climate domain, the dominant covariate differs across algorithms, with thermal radiation most prominent for the GeoEvolve-evolved version.

\begin{figure}[H]
 \centering
 \includegraphics[width=\linewidth]{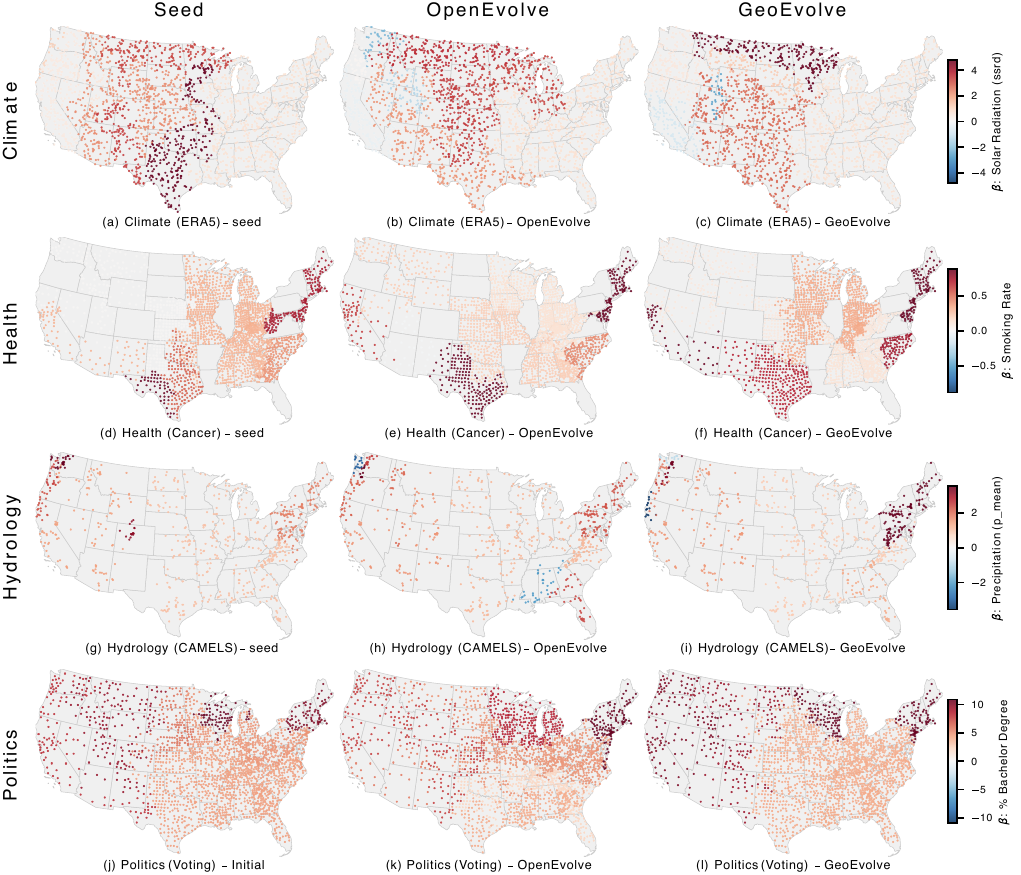}
 \caption{Spatial heterogeneity of one selected regression coefficient across regimes and algorithm variants.}
 \label{fig:coefficient_maps}
\end{figure}

\paragraph{Spatially varying regression coefficients}
Figure~\ref{fig:coefficient_maps} illustrates the spatial variation of one scientifically relevant regression coefficient per domain: solar radiation ($\beta_7$) for climate, smoking rate ($\beta_5$) for health, precipitation ($\beta_1$) for hydrology, and the percentage of bachelor's degrees ($\beta_4$) for politics. In the climate domain, the solar radiation coefficient is strongly positive in the western US, where solar insolation drives surface temperature, and decreases eastward under the influence of cloud cover and humidity. GeoEvolve produces sharper transitions near the 100th meridian, consistent with the well-documented aridity boundary. The precipitation coefficient in the hydrology domain is strongly positive in high-runoff regimes (Pacific Northwest, Appalachian highlands) and weakly positive in arid regimes; this pattern is preserved across algorithms but is more sharply drawn by GeoEvolve.

\baselineskip12pt
\bibliographystyle{elsarticle-harv}
\bibliography{02_references}

@article{guo2023extending,
  title={Extending regionalization algorithms to explore spatial process heterogeneity},
  author={Guo, Hao and Python, Andre and Liu, Yu},
  journal={International Journal of Geographical Information Science},
  volume={37},
  number={11},
  pages={2319--2344},
  year={2023},
  publisher={Taylor \& Francis}
}

@article{tobler1970computer,
  title={A computer movie simulating urban growth in the {Detroit} region},
  author={Tobler, Waldo R},
  journal={Economic Geography},
  volume={46},
  number={sup1},
  pages={234--240},
  year={1970},
  publisher={Taylor \& Francis}
}

@article{li2023autonomous,
  title={Autonomous {GIS}: the next-generation {AI}-powered {GIS}},
  author={Li, Zhenlong and Ning, Huan},
  journal={International Journal of Digital Earth},
  volume={16},
  number={2},
  pages={4668--4686},
  year={2023},
  publisher={Taylor \& Francis}
}

@article{zhang2023geogpt,
  title={{GeoGPT}: Understanding and processing geospatial tasks through an autonomous {GPT}},
  author={Zhang, Yifan and Wei, Cheng and Wu, Shangyou and He, Zhengting and Yu, Wenhao},
  journal={arXiv preprint arXiv:2307.07930},
  year={2023}
}

@article{ning2025autonomous,
  title={An autonomous {GIS} agent framework for geospatial data retrieval},
  author={Ning, Huan and Li, Zhenlong and Akinboyewa, Temitope and Lessani, M Naser},
  journal={International Journal of Digital Earth},
  volume={18},
  number={1},
  pages={2458688},
  year={2025},
  publisher={Taylor \& Francis}
}

@article{lin2025shapefilegpt,
  title={{ShapefileGPT}: A multi-agent large language model framework for automated shapefile processing},
  author={Lin, Qingming and Hu, Rui and Li, Huaxia and Wu, Sensen and Li, Yadong and Fang, Kai and Feng, Hailin and Du, Zhenhong and Xu, Liuchang},
  journal={International Journal of Digital Earth},
  volume={18},
  number={2},
  pages={2577884},
  year={2025},
  publisher={Taylor \& Francis}
}

@article{zhang2024mapgpt,
  title={{MapGPT}: an autonomous framework for mapping by integrating large language model and cartographic tools},
  author={Zhang, Yifan and He, Zhengting and Li, Jingxuan and Lin, Jianfeng and Guan, Qingfeng and Yu, Wenhao},
  journal={Cartography and Geographic Information Science},
  volume={51},
  number={6},
  pages={717--743},
  year={2024},
  publisher={Taylor \& Francis}
}

@article{meyerson2024language,
  title={Language model crossover: Variation through few-shot prompting},
  author={Meyerson, Elliot and Nelson, Mark J and Bradley, Herbie and Gaier, Adam and Moradi, Arash and Hoover, Amy K and Lehman, Joel},
  journal={ACM Transactions on Evolutionary Learning},
  volume={4},
  number={4},
  pages={1--40},
  year={2024},
  publisher={ACM New York, NY}
}

@article{ma2023eureka,
  title={Eureka: Human-level reward design via coding large language models},
  author={Ma, Yecheng Jason and Liang, William and Wang, Guanzhi and Huang, De-An and Bastani, Osbert and Jayaraman, Dinesh and Zhu, Yuke and Fan, Linxi and Anandkumar, Anima},
  journal={arXiv preprint arXiv:2310.12931},
  year={2023}
}

@article{romera2024mathematical,
  title={Mathematical discoveries from program search with large language models},
  author={Romera-Paredes, Bernardino and Barekatain, Mohammadamin and Novikov, Alexander and Balog, Matej and Kumar, M Pawan and Dupont, Emilien and Ruiz, Francisco JR and Ellenberg, Jordan S and Wang, Pengming and Fawzi, Omar and others},
  journal={Nature},
  volume={625},
  number={7995},
  pages={468--475},
  year={2024},
  publisher={Nature Publishing Group UK London}
}

@article{shojaee2024llm,
  title={{LLM-SR}: Scientific equation discovery via programming with large language models},
  author={Shojaee, Parshin and Meidani, Kazem and Gupta, Shashank and Farimani, Amir Barati and Reddy, Chandan K},
  journal={arXiv preprint arXiv:2404.18400},
  year={2024}
}

@article{novikov2025alphaevolve,
  title={{AlphaEvolve}: A coding agent for scientific and algorithmic discovery},
  author={Novikov, Alexander and V{\~u}, Ng{\^a}n and Eisenberger, Marvin and Dupont, Emilien and Huang, Po-Sen and Wagner, Adam Zsolt and Shirobokov, Sergey and Kozlovskii, Borislav and Ruiz, Francisco JR and Mehrabian, Abbas and others},
  journal={arXiv preprint arXiv:2506.13131},
  year={2025}
}

@article{luo2025geoevolve,
  title={{GeoEvolve}: Automating Geospatial Model Discovery via Multi-Agent Large Language Models},
  author={Luo, Peng and Lou, Xiayin and Zheng, Yu and Zheng, Zhuo and Ermon, Stefano},
  journal={arXiv preprint arXiv:2509.21593},
  year={2025}
}

@misc{openevolve,
  title = {{OpenEvolve}: an open-source evolutionary coding agent},
  author = {Asankhaya Sharma},
  year = {2025},
  howpublished = {GitHub repository},
  url = {https://github.com/algorithmicsuperintelligence/openevolve}
}

@article{anselin2024endogenous,
  title={Endogenous spatial regimes},
  author={Anselin, Luc and Amaral, Pedro},
  journal={Journal of Geographical Systems},
  volume={26},
  number={2},
  pages={209--234},
  year={2024},
  publisher={Springer}
}

@article{brunsdon1998geographically,
  title={Geographically weighted regression},
  author={Brunsdon, Chris and Fotheringham, Stewart and Charlton, Martin},
  journal={Journal of the Royal Statistical Society: Series D (The Statistician)},
  volume={47},
  number={3},
  pages={431--443},
  year={1998},
  publisher={Wiley Online Library}
}

@article{lou2025geoconformal,
  title={{GeoConformal} prediction: a model-agnostic framework for measuring the uncertainty of spatial prediction},
  author={Lou, Xiayin and Luo, Peng and Meng, Liqiu},
  journal={Annals of the American Association of Geographers},
  volume={115},
  number={8},
  pages={1971--1998},
  year={2025},
  publisher={Taylor \& Francis}
}

@article{singh2025openai,
  title={{OpenAI GPT-5} system card},
  author={Singh, Aaditya and Fry, Adam and Perelman, Adam and Tart, Adam and Ganesh, Adi and El-Kishky, Ahmed and McLaughlin, Aidan and Low, Aiden and Ostrow, AJ and Ananthram, Akhila and others},
  journal={arXiv preprint arXiv:2601.03267},
  year={2025}
}

@article{team2024gemini,
  title={{Gemini}: A Family of Highly Capable Multimodal Models},
  author={{Gemini Team} and Anil, Rohan and Borgeaud, Sebastian and Alayrac, Jean-Baptiste and Yu, Jiahui and Soricut, Radu and Schalkwyk, Johan and Dai, Andrew M and Hauth, Anja and Millican, Katie and others},
  journal={arXiv preprint arXiv:2312.11805},
  year={2023}
}

@article{yang2025qwen3,
  title={Qwen3 technical report},
  author={Yang, An and Li, Anfeng and Yang, Baosong and Zhang, Beichen and Hui, Binyuan and Zheng, Bo and Yu, Bowen and Gao, Chang and Huang, Chengen and Lv, Chenxu and others},
  journal={arXiv preprint arXiv:2505.09388},
  year={2025}
}

@article{munoz2021era5,
  title={{ERA5-Land}: A state-of-the-art global reanalysis dataset for land applications},
  author={Mu{\~n}oz-Sabater, Joaqu{\'\i}n and Dutra, Emanuel and Agust{\'\i}-Panareda, Anna and Albergel, Cl{\'e}ment and Arduini, Gabriele and Balsamo, Gianpaolo and Boussetta, Souhail and Choulga, Margarita and Harrigan, Shaun and Hersbach, Hans and others},
  journal={Earth System Science Data},
  volume={13},
  number={9},
  pages={4349--4383},
  year={2021},
  publisher={Copernicus GmbH}
}

@article{addor2017camels,
  title={The CAMELS data set: catchment attributes and meteorology for large-sample studies},
  author={Addor, Nans and Newman, Andrew J and Mizukami, Naoki and Clark, Martyn P},
  journal={Hydrology and Earth System Sciences},
  volume={21},
  number={10},
  pages={5293--5313},
  year={2017},
  publisher={Copernicus Publications G{\"o}ttingen, Germany}
}

@article{fotheringham2023measuring,
  title={Measuring the unmeasurable: models of geographical context},
  author={Fotheringham, A Stewart and Li, Ziqi},
  journal={Annals of the American Association of Geographers},
  volume={113},
  number={10},
  pages={2269--2286},
  year={2023},
  publisher={Taylor \& Francis}
}

@article{greenlund2022places,
  title={{PLACES}: local data for better health},
  author={Greenlund, Kurt J and Lu, Hua and Wang, Yan and Matthews, Kevin A and LeClercq, Jennifer M and Lee, Benjamin and Carlson, Susan A},
  journal={Preventing Chronic Disease},
  volume={19},
  pages={E31},
  year={2022}
}

@book{fotheringham2002geographically,
  title={Geographically weighted regression: the analysis of spatially varying relationships},
  author={Fotheringham, A Stewart and Brunsdon, Chris and Charlton, Martin},
  year={2002},
  publisher={John Wiley \& Sons}
}

@article{fotheringham2017multiscale,
  title={Multiscale geographically weighted regression ({MGWR})},
  author={Fotheringham, A Stewart and Yang, Wenbai and Kang, Wei},
  journal={Annals of the American Association of Geographers},
  volume={107},
  number={6},
  pages={1247--1265},
  year={2017},
  publisher={Taylor \& Francis}
}

@book{anselin1988spatial,
  title={Spatial econometrics: methods and models},
  author={Anselin, Luc},
  year={1988},
  publisher={Springer Science \& Business Media}
}

@article{anselin1995local,
  title={Local indicators of spatial association---{LISA}},
  author={Anselin, Luc},
  journal={Geographical Analysis},
  volume={27},
  number={2},
  pages={93--115},
  year={1995},
  publisher={Wiley Online Library}
}

@article{goodchild2004validity,
  title={The validity and usefulness of laws in geographic information science and geography},
  author={Goodchild, Michael F},
  journal={Annals of the Association of American Geographers},
  volume={94},
  number={2},
  pages={300--303},
  year={2004},
  publisher={Taylor \& Francis}
}

@article{janowicz2020geoai,
  title={{GeoAI}: spatially explicit artificial intelligence techniques for geographic knowledge discovery and beyond},
  author={Janowicz, Krzysztof and Gao, Song and McKenzie, Grant and Hu, Yingjie and Bhaduri, Budhendra},
  journal={International Journal of Geographical Information Science},
  volume={34},
  number={4},
  pages={625--636},
  year={2020},
  publisher={Taylor \& Francis}
}

@article{duque2007supervised,
  title={Supervised regionalization methods: A survey},
  author={Duque, Juan C and Ramos, Raul and Suri{\~n}ach, Jordi},
  journal={International Regional Science Review},
  volume={30},
  number={3},
  pages={195--220},
  year={2007},
  publisher={Sage Publications}
}

@article{assuncao2006efficient,
  title={Efficient regionalization techniques for socio-economic geographical units using minimum spanning trees},
  author={Assun{\c{c}}{\~a}o, Renato M and Neves, Marcos Corr{\^e}a and C{\^a}mara, Gilberto and da Costa Freitas, Corina},
  journal={International Journal of Geographical Information Science},
  volume={20},
  number={7},
  pages={797--811},
  year={2006},
  publisher={Taylor \& Francis}
}

@book{lesage2009introduction,
  title={Introduction to spatial econometrics},
  author={LeSage, James and Pace, Robert Kelley},
  year={2009},
  publisher={Chapman and Hall/CRC}
}

@book{cressie1993statistics,
  title={Statistics for spatial data},
  author={Cressie, Noel},
  year={1993},
  publisher={John Wiley \& Sons}
}

@article{mai2022opportunities,
  title={On the Opportunities and Challenges of Foundation Models for Geospatial Artificial Intelligence},
  author={Mai, Gengchen and Huang, Weiming and Sun, Jin and Song, Suhang and Mishra, Deepak and Liu, Ninghao and Gao, Song and Liu, Tianming and Cong, Gao and Hu, Yingjie and others},
  journal={arXiv preprint arXiv:2304.06798},
  year={2023}
}

@article{park2023generative,
  title={Generative agents: Interactive simulacra of human behavior},
  author={Park, Joon Sung and O'Brien, Joseph C and Cai, Carrie J and Morris, Meredith Ringel and Liang, Percy and Bernstein, Michael S},
  journal={Proceedings of the 36th Annual ACM Symposium on User Interface Software and Technology},
  pages={1--22},
  year={2023}
}

@article{liu2024deepseek,
  title={{DeepSeek-V3} technical report},
  author={Liu, Aixin and Feng, Bei and Xue, Bing and Wang, Bingxuan and Wu, Bochao and Lu, Chengda and Zhao, Chenggang and Deng, Chengqi and Zhang, Chenyu and Ruan, Chong and others},
  journal={arXiv preprint arXiv:2412.19437},
  year={2024}
}

@article{openshaw1977geographical,
  title={A geographical solution to scale and aggregation problems in region-building, partitioning and spatial modelling},
  author={Openshaw, Stan},
  journal={Transactions of the Institute of British Geographers},
  pages={459--472},
  year={1977},
  publisher={JSTOR},
  volume={2},
  number={4}
}

@article{openshaw1995algorithms,
  title={Algorithms for reengineering 1991 {Census} geography},
  author={Openshaw, Stan and Rao, Liang},
  journal={Environment and Planning A},
  volume={27},
  number={3},
  pages={425--446},
  year={1995},
  publisher={SAGE Publications Sage UK: London, England}
}

@article{helbich2013data,
  title={Data-driven regionalization of housing markets},
  author={Helbich, Marco and Brunauer, Wolfgang and Hagenauer, Julian and Leitner, Michael},
  journal={Annals of the Association of American Geographers},
  volume={103},
  number={4},
  pages={871--889},
  year={2013},
  publisher={Taylor \& Francis}
}

@article{akinboyewa2025gis,
  title={{GIS Copilot}: towards an autonomous {GIS} agent for spatial analysis},
  author={Akinboyewa, Temitope and Li, Zhenlong and Ning, Huan and Lessani, M Naser},
  journal={International Journal of Digital Earth},
  volume={18},
  number={1},
  pages={2497489},
  year={2025},
  publisher={Taylor \& Francis}
}

@article{zhou2026nora,
  title={{NORA}: A Harness-Engineered Autonomous Research Agent for End-to-End Spatial Data Science},
  author={Zhou, Bing and Huang, Xiao and Ning, Huan and Wu, Qiusheng and Li, Diya and Zhang, Ziyi},
  journal={arXiv preprint arXiv:2605.02092},
  year={2026}
}

@article{li2025giscience,
  title={{GIScience} in the era of Artificial Intelligence: a research agenda towards Autonomous {GIS}},
  author={Li, Zhenlong and Ning, Huan and Gao, Song and Janowicz, Krzysztof and Li, Wenwen and Shekhar, Shashi and Hodgson, Michael E and others},
  journal={Annals of GIS},
  volume={31},
  number={4},
  pages={501--536},
  year={2025},
  publisher={Taylor \& Francis}
}

@article{li2026multiagent,
  title={Discovering Multiagent Learning Algorithms with Large Language Models},
  author={Li, Zun and Schultz, John and Hennes, Daniel and Lanctot, Marc},
  journal={arXiv preprint arXiv:2602.16928},
  year={2026}
}

@inproceedings{deb2000nsga,
  title={A fast elitist non-dominated sorting genetic algorithm for multi-objective optimization: {NSGA-II}},
  author={Deb, Kalyanmoy and Agrawal, Samir and Pratap, Amrit and Meyarivan, Thirunavukkarasu},
  booktitle={International Conference on Parallel Problem Solving from Nature},
  pages={849--858},
  year={2000},
  organization={Springer}
}

@article{mouret2015illuminating,
  title={Illuminating search spaces by mapping elites},
  author={Mouret, Jean-Baptiste and Clune, Jeff},
  journal={arXiv preprint arXiv:1504.04909},
  year={2015}
}

@article{chen2026ethical,
  title={Ethical Spatial Analysis: Addressing Endogenous Bias Through Visual Analytics},
  author={Chen, Chuan and Luo, Peng and Zhao, Bo and Wang, Shengkai and Feng, Yu and Meng, Liqiu},
  journal={Journal of Geovisualization and Spatial Analysis},
  volume={10},
  number={1},
  pages={4},
  year={2026},
  doi={10.1007/s41651-025-00243-2}
}


\end{document}